 \documentclass[twocolumn]{aastex631}
\usepackage{amsmath}
\usepackage{breqn}

\font\manual=manfnt at 7pt \def\dbend{\hbox{\raise0.9ex\hbox{\manual\char127\hspace{0.6em}}}}

\providecommand{\e}[1]{\ensuremath{\times 10^{#1}}}

\newcounter{INTERNALionstage}

\def\gtsim{\mathrel{\hbox{\rlap{\hbox{\lower4pt\hbox{$\sim$}}}\hbox{$>$}}}}
\def\lesssim{\mathrel{\hbox{\rlap{\hbox{\lower4pt\hbox{$\sim$}}}\hbox{$<$}}}}

\def\A{{\rm\thinspace \AA}}

\def\g{{\rm\thinspace g}}

\def\K{{\rm\thinspace K}}

\def\m{{\rm\thinspace m}}

\def\s{{\rm\thinspace s}}

\def\W{{\rm\thinspace W}}

\def\h0{\mbox{{\rm H}$^0$}}
\DeclareMathAlphabet{\vib}{OML}{cmm}{m}{it}

\renewcommand{\thefootnote}{\fnsymbol{footnote}}

\shorttitle{Correlated variability between AMS and 
Chandra/XMM-Newton}
\graphicspath{{./}{figures/}}

\begin{document}

\title{
Advancing Precision Particle Background Estimation for Future X-ray Missions: Correlated Variability between AMS and Chandra/XMM-Newton}

\author[0000-0002-5222-1337]{Arnab Sarkar}
\affiliation{MIT Kavli Institute for Astrophysics and Space Research, 70 Vassar St, Cambridge, MA 02139, USA}
\email{arnabsar@mit.edu}

\author[0000-0002-4737-1373]{Catherine E.\ Grant}
\affiliation{MIT Kavli Institute for Astrophysics and Space Research, 70 
Vassar St, Cambridge, MA 02139, USA}

\author[0000-0002-3031-2326]{Eric D.\ Miller}
\affiliation{MIT Kavli Institute for Astrophysics and Space Research, 70 
Vassar St, Cambridge, MA 02139, USA}

\author[ 0000-0002-1379-4482]{Mark Bautz}
\affiliation{MIT Kavli Institute for Astrophysics and Space Research, 70 
Vassar St, Cambridge, MA 02139, USA}

\author[0000-0003-4876-7756]{Benjamin Schneider}
\affiliation{MIT Kavli Institute for Astrophysics and Space Research, 70 
Vassar St, Cambridge, MA 02139, USA}

\author{Rick F. Foster}
\affiliation{MIT Kavli Institute for Astrophysics and Space Research, 70 
Vassar St, Cambridge, MA 02139, USA}

\author[0000-0002-4962-0740]{Gerrit Schellenberger}
\affiliation{Center for Astrophysics $\vert$ Harvard \& Smithsonian, 
60 Garden st, Cambridge, MA 02139, USA}

\author[0000-0003-0667-5941]{Steven Allen}
\affiliation{Stanford University,
382 Via Pueblo Mall,
Stanford,  California 94305, USA}

\author[0000-0002-0765-0511]{Ralph P. Kraft}
\affiliation{Center for Astrophysics $\vert$ Harvard \& Smithsonian, 
60 Garden st, Cambridge, MA 02139, USA}

\author[0000-0003-0667-5941]{Dan Wilkins}
\affiliation{Stanford University,
382 Via Pueblo Mall,
Stanford,  California 94305, USA}

\author{Abe Falcone}
\affiliation{
Pennsylvania State University,
0525 Davey Laboratory, University Park, PA 16802, USA}

\author{Andrew Ptak}
\affiliation{
NASA Goddard Space Flight Center, Greenbelt, MD20771, USA}



\begin{abstract}

Galactic cosmic ray (GCR)
particles have a significant 
impact on the particle-induced 
background of X-ray
observatories, and their
flux exhibits substantial
temporal variability, 
potentially influencing 
background levels.
In this study, we present
one-day binned high-energy
reject rates derived from
the Chandra-ACIS and XMM-Newton
EPIC-pn instruments, 
serving as proxies for
GCR particle flux. 
We systematically analyze 
the ACIS and EPIC-pn reject 
rates and compare them with
the AMS proton flux. 
Our analysis initially 
reveals robust correlations
between the AMS proton flux and 
the ACIS/EPIC-pn reject rates
when binned over 27-day intervals. 
However, a closer examination 
reveals substantial fluctuations 
within each 27-day bin, 
indicating shorter-term 
variability.
Upon daily binning, we observe 
finer temporal structures in
the datasets, demonstrating
the presence of recurrent 
variations with periods of 
$\sim$ 25 days and
23 days in ACIS and EPIC-pn 
reject rates, respectively,
spanning the years 2014 to 2018. 
Notably, during the 2016–2017 
period, 
we additionally detect 
periodicities of $\sim$~13.5 days 
and 9 days 
in the ACIS and EPIC-pn reject 
rates, 
respectively. 
Intriguingly, we observe a 
time lag of $\sim$ 6
days between the AMS proton flux 
and the ACIS/EPIC-pn reject 
rates during the second half
of 2016. 
This time lag is not visible before 2016 and after
2017. 
The underlying
physical mechanisms responsible
for this time lag remain a 
subject of ongoing 
investigation.

\end{abstract}




\section{Introduction}\label{sec:intro}
Athena (Advanced Telescope for 
High ENergy Astrophysics), 
ESA’s next large X-ray observatory,
is scheduled to launch in 
the mid 2030s to Earth-Sun L2
\citep{2013arXiv1306.2307N}.
Some of Athena's 
primary scientific objectives 
align with the study of 
faint diffuse emission, 
including those from galaxy 
clusters, groups of galaxies, and 
the intergalactic medium (IGM). 
These are also primary science
goals of 
current (e.g., Chandra, XMM-Newton, 
eROSITA, XRISM) and 
future  
(Lynx; 
\citealt{2019JATIS...5b1001G},
LEM; \citealt{2022arXiv221109827K},
AXIS; \citealt{Reynoldsetal2023}) X-ray telescopes.
Achieving these objectives 
hinges on precise control 
of the total flux and 
minimizing systematic
uncertainties linked to our 
understanding of the
non-X-ray background
\citep{2021MNRAS.501.3767S,2022MNRAS.516.3068S,2022ApJ...935L..23S,2023ApJ...944..132S,2024ApJ...962..161S}.
Extensive efforts have 
been invested in 
understanding, modeling, 
and mitigating this 
background component 
originating from particles and 
high-energy photons, 
{ for example, the Self-Anti-
Coincidence (SAC)
technique, \citealt{2022JATIS...8a8001M},
machine learning approach,
\citealt{10.1117/12.2677095},
and many others}
\citep[e.g.,][]{2018SPIE10699E..4HG,2020ApJ...891...13B,2022SPIE12181E..2EG,2022ApJ...928..168G}.
The non-X-ray background 
has consistently posed a 
significant challenge in
X-ray missions, limiting 
the full exploitation of 
scientific data from diffuse 
sources
\citep{2017ExA....44..263M}.

The non-X-ray background 
arises primarily from two
sources: soft protons 
focused onto the focal plane
by the telescope mirrors 
and unfocused background 
generated by various particles
\citep{2022hxga.book...39C}.
Soft protons, originating in 
the solar corona and Earth's 
magnetosphere, 
possess energies typically 
below a few 100 keV and 
can reach the detectors due 
to their concentration by 
the telescope mirrors.
The spectral shape of this
component can be approximated 
by a power-law continuum, 
characterized by highly 
variable intensity and slope
\citep{2008A&A...478..575K}. 
Soft protons, when present,
can dramatically elevate the
overall background intensity 
in brief time 
intervals,
often referred to as 
``soft proton flares"
\citep{2008A&A...478..575K}.
These protons primarily
deposit their energy near 
the detector's surface and
generate event patterns
identical to valid X-ray 
photons, making them challenging 
to distinguish and reject
during onboard processing 
\citep{2017ExA....44..321G}.

The unfocused particle-induced 
background can arise from several 
sources, including  Galactic cosmic 
rays (GCRs), 
solar energetic protons, 
and trapped radiation within
Earth's magnetosphere. 
The characteristics of this 
background, including its
temporal behavior, spectral
features, and spatial
distribution, depend on the 
primary particle's energy. 
GCR particles, typically
with energies greater than
10 MeV and 
composed mainly of protons, 
electrons, 
and helium ions, 
exhibit variability 
linked to the solar cycle.  
These incoming particles 
interact with the detector and 
surrounding components,
giving rise to secondary
particles
\citep[e.g.,][]{2008A&A...478..575K,2018SPIE10699E..1IV}. 
To prevent these 
high-energy particles
from saturating 
the limited bandwidth for 
telemetry,
onboard event processing is 
employed to reject events
generated by primary GCRs, 
primarily due to their high 
total energies or their
distinctive pixel patterns
\citep{2002A&A...389...93L}. 
However, secondary 
electrons and photons 
resulting from these 
unfocused components deposit 
charge in the detector, 
posing challenges in 
distinguishing them from X-ray 
events originating from 
celestial sources.
Consequently, they
significantly contribute to, 
and often dominate, 
the residual (unrejected) 
instrumental background.

Quantifying the 
particle-induced
background of X-ray
observatories is a complex 
task due to its pronounced 
temporal variability, 
further complicated by the
presence or absence of 
the geomagnetic shielding.
The transportation of 
GCRs within the heliosphere
is governed
by the interplay 
between the solar wind and 
the regular as well as 
turbulent components of
the heliospheric magnetic
field (HMF). 
As GCRs interact with the 
solar wind and the HMF,
their spectrum below a
few dozen GV
undergoes modulation 
relative to the local 
interstellar spectrum  
(LIS; \citealt{2013LRSP...10....3P}). 
This modulation of
GCR flux leads
to a range of { quasi-periodic} 
variations 
occurring on various timescales, 
spanning from hours to several 
years. 
These phenomena have been 
extensively studied in
\citet{2017LRSP...14....3U,2016ScChE..59..547K,2016SoPh..291..581C,2014SSRv..186..359B,2021A&A...646A.128M} 
and references therein.

GCR particles are routinely 
detected and characterized
using the Alpha Magnetic 
Spectrometer (AMS-02)
instrument, a specialized 
particle detector installed 
on the International Space Station 
(ISS) 
\citep{2012IJMPE..2130005K,2015PhRvL.114q1103A}. 
AMS provides precise, synoptic
measurements of GCR particle
fluxes and energies on timescales 
as short as 1 day 
\citep{2021PhRvL.127A1102A}. 
These data, when compared to
background rates measured aboard
X-ray observatories can provide 
insight into the mechanisms
responsible for level and 
variability of the 
particle-induced  background 
 in X-ray instruments.  

When a GCR particle interacts 
with an X-ray detector, 
it leaves behind a 
distinctive particle track
marked by multiple 
pixels { with large signal charge}.
Various X-ray telescopes
employ specific algorithms to 
identify such invalid (i.e. non-X-ray) events.

These algorithms exploit the fact 
that X-ray optics focus incident 
cosmic X-rays over a limited 
spectral band with a
well-defined upper energy limit.
Thus, for instance, the ACIS 
instrument onboard Chandra
has an on-board filter that rejects events with energies surpassing a 
designated threshold.
Similarly, the EPIC-pn 
instrument onboard XMM-Newton 
discards entire columns of 
pixels that contain 
pixels with
energies exceeding a
predefined threshold.
Although these detectors
do not directly measure 
GCR particle fluxes, 
the count of discarded events, 
expressed as reject rates 
per frame per frame time,
can serve as a proxy measurement 
of GCR particle fluxes, 
as shown in several
previous studies 
\citep{2018SPIE10699E..4HG,2020ApJ...891...13B,2022ApJ...928..168G}.
Notably, \citet{2022SPIE12181E..2EG} 
demonstrated a strong 
correlation between AMS proton 
fluxes and ACIS reject
rates binned over 27--day 
intervals, further
affirming the efficacy
of X-ray reject rates 
as proxies for GCR
particle fluxes.

Despite the critical 
relevance of the
variabilities seen in the GCR 
particle flux 
\citep{2021PhRvL.127A1102A}, 
their impact on the variability of the
particle 
background as seen 
by X-ray detectors has been 
relatively unexplored, 
primarily due to the 
limited availability 
of non-X-ray-background 
information over long 
intervals at fine time bin.
In this paper, for the
first time
we present results 
derived from the analysis 
of one-day binned ACIS and 
EPIC-pn reject rates, 
encompassing data from
nearly all archival science 
observations spanning the
years 2011--2020. 
We examine the long-term
and short-term variabilities 
in both datasets and compare
them with that of AMS proton
fluxes. 
We discuss how this
knowledge can inform
future missions, by inclusion
of special purpose particle
monitors, or by 
comparison to other space-based 
particle monitors.
The overarching objective
of this work is to 
enhance our understanding of 
instrumental 
background characterization 
and mitigation strategies, 
particularly relevant to 
the Athena Wide-Field Imager (WFI). 
Results of this
work, however, can be extended 
and
applied to any future
X-ray missions.

\section{Data gathering}\label{sec:data}
GCR protons are 
one of the important
sources of quiescent 
particle-background, 
modulated by the solar cycle 
and solar activity.
We analyze 9 years (2011--2020) of 
archival
data on proton flux and high-energy 
reject rates from AMS, Chandra, and
XMM-Newton.

\subsection{Alpha Magnetic Spectrometer (AMS-02)}\label{ssec:ams}
The { 
AMS-02 (hereafter AMS)
} is 
a state-of-the-art particle
physics experiment module installed on the 
ISS
in May 2011. 
Developed to explore the cosmos for
high-energy cosmic rays and particles, 
the AMS plays a crucial role in advancing our 
understanding of fundamental physics phenomena. 
This sophisticated instrument is equipped with
permanent magnet,
silicon tracker and 
time of flight (TOF) scintillation counters
that allow it to 
identify and analyze the properties of 
cosmic particles with 
unparalleled precision 
\citep{2015PhRvL.114q1103A}.
For the detailed layout of the detector, we
refer readers to \citet{2012IJMPE..2130005K}.
The large acceptance and high 
precision of the AMS allow it to perform 
accurate measurements of the 
particle fluxes as functions of time and energy.
Its unique vantage point in
space provides data that cannot be 
easily obtained from Earth-based experiments,
providing
unique opportunities to
probe the dynamics of solar modulation
\citep{2017arXiv171203178T}, 
and the 
GCR propagation
\citep{2017ApJ...840..115B}.

Though AMS orbits in Low Earth Orbit (LEO), 
proton flux measured by AMS can be 
reconstructed to anticipate the flux 
that would occur in the 
absence of Earth's magnetic field 
shielding
\citep{2012IJMPE..2130005K,2015PhRvL.114q1103A}.
{ By understanding the ISS's
path around the Earth,
the observed cosmic ray spectrum 
can be corrected for the effects
of geomagnetic shielding at each 
point in the ISS orbit to obtain
the spectrum incident outside 
the Earth's magnetosphere
\citep{2015PhRvL.114q1103A}
}
Since the ISS's orbital 
journey exposes it to varying 
levels of cut-off 
rigidity, the GCR proton spectrum
can be
reconstructed as without
the influence of 
Earth's magnetic field. 
The AMS team has 
successfully accomplished 
this reconstruction, 
including assessments of the
potential errors 
introduced by this process,
as well as by the 
other calibration and filtering
procedures 
and made this data available
to the wider community
\citep{2015PhRvL.114q1103A,2021PhRvL.127A1102A}.

Regarding the GCR 
proton flux, there have been
three significant
data releases in the literature 
by the AMS team.
Firstly, the precise 
measurement of the proton spectrum 
conducted by \citet{2015PhRvL.114q1103A}. 
Following that, an investigation
into the temporal patterns of proton 
fluxes on timescales of the Bartels rotation period
(a period of 27 days, approximately 
equivalent to the synodic solar
rotation period), 
as detailed in \citet{2018PhRvL.121e1101A}. 
Lastly, \citet{2021PhRvL.127A1102A} 
presents the daily variations in proton 
fluxes. 
In the scope of our current research,
we employ these meticulously calibrated daily 
proton 
flux data spanning from 2011 to 2019. 
For this work, we summed all the proton
fluxes in the 0.4--60 GeV energy band
for a specific time bin.
This dataset enables us to 
examine 
the temporal intricacies of GCR protons 
while 
considering the particle background effects 
relevant to future X-ray telescopes.

\subsection{Chandra ACIS}\label{ssec:acis}
The Advanced CCD Imaging Spectrometer 
(ACIS)
onboard the Chandra X-ray Observatory
was launched
into a highly elliptical 2.7 day
orbit in 1999.
ACIS employs an assembly of
framestore charge-coupled devices 
(CCDs), 
comprising a total of ten CCDs. 
Each individual CCD encompasses 
a matrix of 1024 by 1024 pixels,
each with dimensions of 24 microns. 
Our particular focus centers on the 
properties of the back-illuminated 
(BI) CCD, 
specifically ACIS-S3, situated 
at the ACIS-S 
aimpoint \citep{2022SPIE12181E..2EG}. 
This choice is motivated by its 
intrinsic resemblance to forthcoming
detectors such as the Athena-WFI.
ACIS-S3 measures 45 microns in thickness, 
operates in a fully depleted state, and is 
configured for photon-counting mode, 
characterized by frametimes of 
approximately 3 
seconds. For detailed layout we refer
readers to \citet{2000SPIE.4012....2W}.

Chandra is in a highly elliptical
orbit (perigee $\sim$ 10,000 km; 
apogee $\sim$ 130,000~km) 
which traverses diverse
particle environments during its 
orbital trajectory, dipping into the Earth's radiation belts around perigee.
Scientific observations, however, all occur well outside Earth's radiation belts and magnetic shielding { at altitudes above roughly 60,000~km 
\citep{2002ASPC..262..401G}.  The Chandra orbit evolves with time with apogee ranging from about 130,000 to 145,000~km during this time period.} 
It is anticipated that the external particle environment encountered by
potential upcoming missions at the Langrage points L1/L2 or in other high-Earth orbits 
should
closely resemble that observed by 
Chandra. 
This similarity underscores the 
significance of ACIS as a 
valuable tool for 
investigating high-energy GCR
particles and comprehending their 
impact on particle background dynamics. 

The detection methodology involves an 
on-board filtering mechanism: events with 
pulse-heights 
surpassing a tunable upper threshold
(generally either 13 keV or 15 keV),
as well as events exhibiting morphologies 
characterized by an abundance of contiguous 
pixels, are systematically excluded.
While the filtered events 
are not telemetered to the ground, 
the instrument maintains a comprehensive
record of the count of eliminated events
and the rationale behind their removal. 
Among these counters, the parameter
denoted as the ``high-energy reject rate" 
serves as a proxy
\footnote{One particle interaction can
potentially create many associated
events in the CCD, 
so there is not a one-to-one 
correspondence between particle rate 
and reject rate, 
but the two quantities should
be related.}
for the
particle rate encountered by the 
spacecraft.

In this study, we investigate high-energy 
reject rates
(reject rates hereafter)
from the period spanning 2011 to 2020. 
These rates are available in the 
standard data products provided with
each ACIS observation.
\footnote{The DROP\_AMP column of the 
level 1 exposure 
statistics file, ``acisf$\star$stat1.fits'',
found in the ``secondary'' 
directory (\url{https://cxc.cfa.harvard.edu/ciao/threads/intro\_data/}).}
In this particular study, 
we made use of a separate
telemetry archive created by 
the ACIS instrument team, 
as a quicker method to access 
the reject rates.
To derive the reject rates,
we employ two distinct datasets. 

\begin{figure}
    \centering
    \begin{tabular}{cc}
         \includegraphics[width=0.5\textwidth]{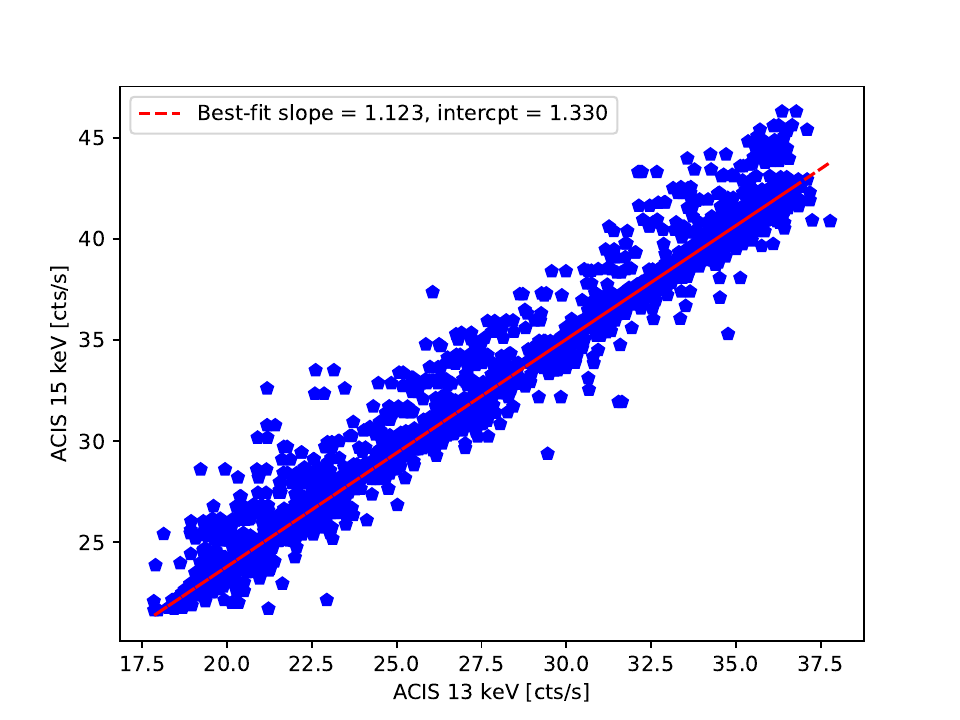}\\
    \end{tabular}
\caption{ Correlation 
between the reject rates
for the 15 keV and 
13 keV pulseheight 
filters of ACIS data. 
Red line shows the best-fit linear
regression.
}
\label{fig:13kev_15keV}
\end{figure}

One set of data with a threshold of
3750 ADU ($\sim$ 15 keV) was collected during instances
when ACIS was in a stowed 
configuration and was exposed to an
external calibration source.
During this configuration, the detector 
remained shielded from focused X-rays. 
Another set of data with a threshold of
3275 ADU ($\sim$ 13 keV) was gathered during the time when
detector
was exposed to the sky
for nominal science operation. 
{
The reject rates were
derived by aligning the 
15 keV dataset with the 
13 keV dataset and combining them.
To achieve this alignment, 
the 15 keV dataset was normalized
using the best-fit parameters 
derived from a linear regression
analysis between the two datasets. 
Figure \ref{fig:13kev_15keV} 
provides a 
visualization 
of these datasets along with their 
corresponding best-fit linear
regression line.
}

It is worth noting that the prevailing quiescent particle 
background primarily arises from GCR
protons and any secondary particles from interactions with the spacecraft and instrument components, whether the instrument is in the stowed position or otherwise. Focused low energy protons ($\sim$ 100 keV) which can cause short term increases in background count rates are not seen while the instrument is stowed and in general produce few events in the reject rate channel. This is further demonstrated by the good agreement in reject rates between the stowed and science observations after scaling.

\subsection{XMM-Newton EPIC-pn}\label{ssec:xmm_pn}
The EPIC-pn CCD camera stands as a key 
instrument on XMM-Newton, featuring a 
collection area of approximately 2500 cm$^2$ at 
1 keV and a field of view spanning 
27.2 $\times$ 26.2 
degrees across the wide energy 
spectrum of 0.1 keV to 12 keV 
\citep{2001A&A...365L..18S}.
{ EPIC-pn consists of 
four individual quadrants (BI) 
each 
having three CCD subunits with 
a 
format 200 $\times$ 64 pixels 
with a pixel size of 150 $\mu$m.}  
Operating in a highly elliptical orbit
like Chandra
{( 
with an apogee of
about 114,000 km
and a perigee of 
about 7000 km
)}, XMM-Newton also offers an 
environment for studying GCR interactions, 
relevant to upcoming X-ray 
telescopes' orbits. 
The significance of XMM-Newton
is twofold: its EPIC-pn camera bears 
similarity to Athena--WFI; 
its elliptical orbit remains unshielded by 
Earth's magnetic field for the majority of its 
trajectory, causing it to encounter the effects 
of high-energy GCRs resembling Athena's L1/L2 
orbit situation
\citep{2021ApJ...908...37M,2020ApJ...891...13B,2022ApJ...928..168G}.

The EPIC background encompasses three
primary components: first, the
background originating from detector
noise and defects; second, the component 
resulting from mirror concentration 
(comprising X-ray photons and low-energy 
protons); 
and third, the unfocused 
high-energy particles 
and their subsequent products
\citep{2004SPIE.5165..112F,2014MNRAS.445.2146F,2021ApJ...908...37M}.
For this work, we primarily focused on the 
unfocused component. 
Certain particles, 
such as MIPs, are identified and 
subsequently discarded onboard the
satellite. 
This leads to time-dependent 
exclusion of CCD 
columns and/or events exhibiting patterns 
inconsistent with valid X-ray events,
primarily due to uneven spatial distribution 
across nearby pixels.

For this present study, we collected
onboard excluded EPIC-pn CCD 
columns due to MIPs,
known as Number of
Discarded Lines (NDSLIN), as a proxy for
high-energy GCR particle rates seen by 
EPIC-pn
\footnote{For detailed description 
on NDSLIN - \url{https://www.cosmos.esa.int/documents/332006/623312/IdlC_NDSLIN_May2019.pdf}}. 
The EPIC-pn data used in this work was taken
when telescope was in normal science 
operation
mode and performed in Full Window Mode (FWM).
We analyze all the archival EPIC-pn data
in FWM mode
spanning 2011--2020. 
Initially, EPIC-pn event files were
generated 
and filtered using
SAS tool -- {\tt epchain}
and {\tt pn-filter}. 
Finally, the NDSLIN rates were derived
from the house-keeping extension contained 
within the EPIC-pn event files.

\section{Results}
We present 
our findings obtained through the
analysis and comparison of
AMS proton flux, 
ACIS high-energy reject rates, 
and PN NDSLIN rates spanning the
period from 2011 to 2020. 

\subsection{Correlation between AMS, ACIS, and
PN}\label{sec:corr}

\begin{figure*}
    \centering
    \begin{tabular}{c}
         \includegraphics[width=0.7\textwidth]{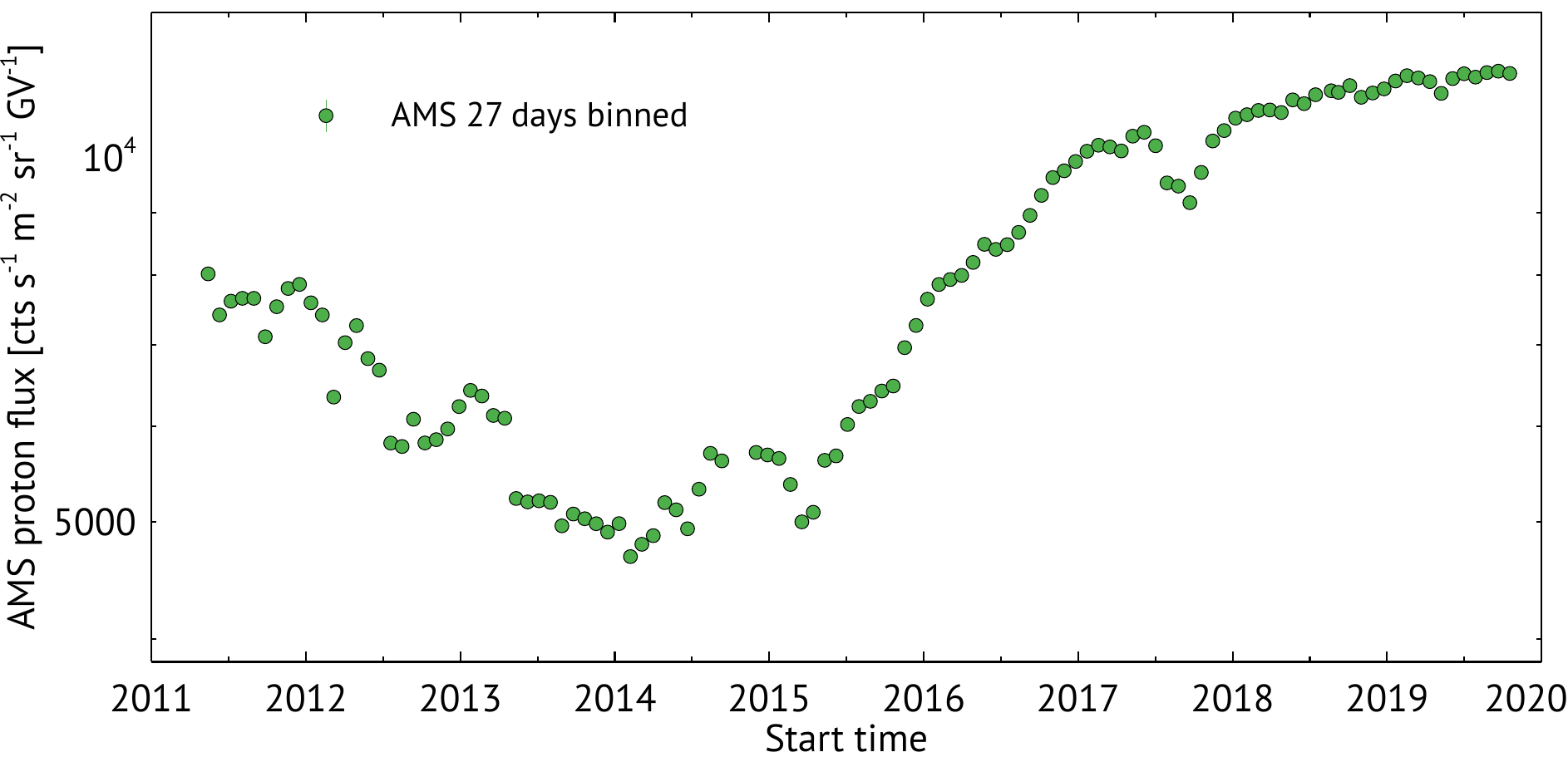}\\  \includegraphics[width=0.7\textwidth]{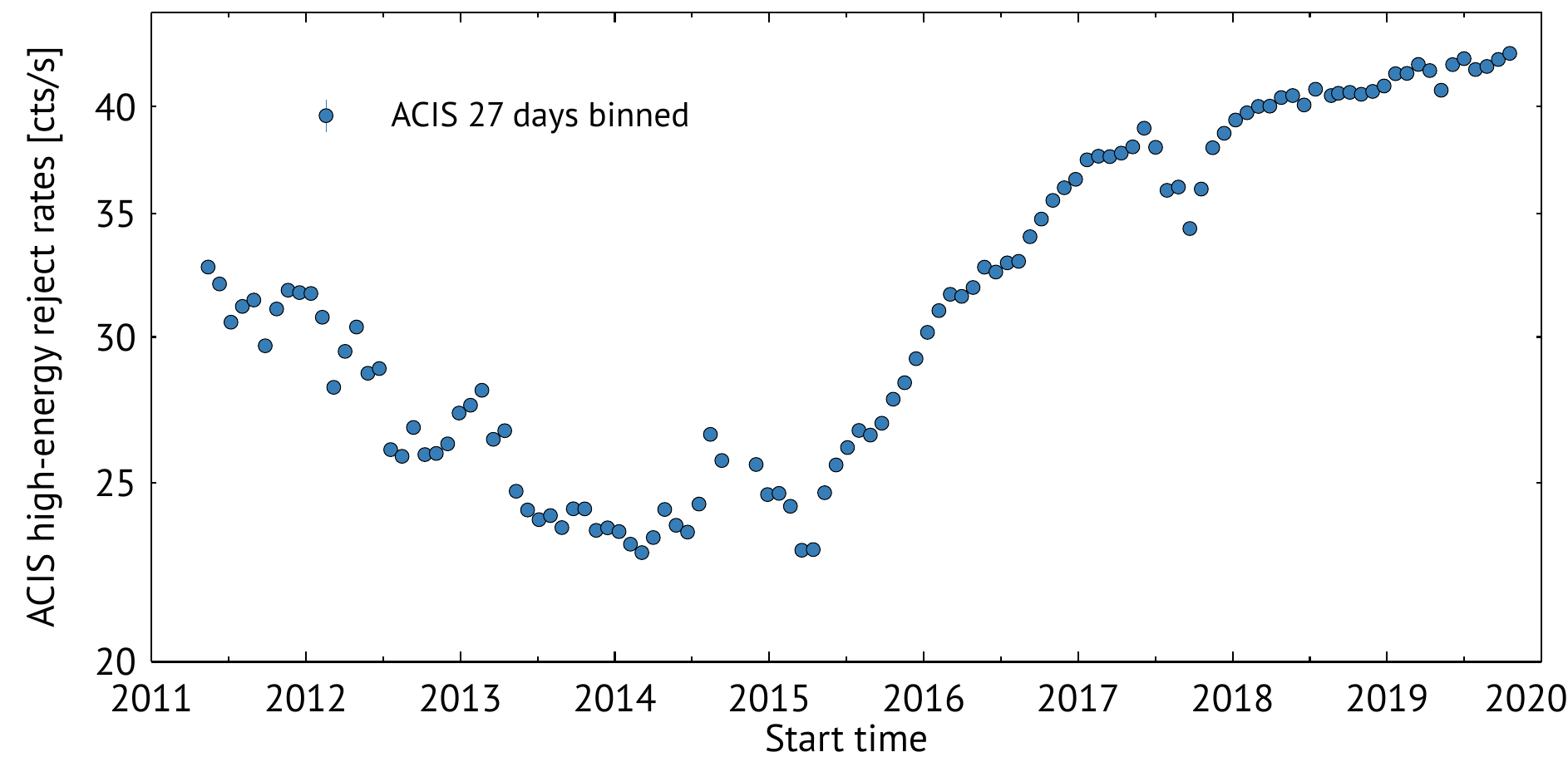}\\
         \includegraphics[width=0.7\textwidth]{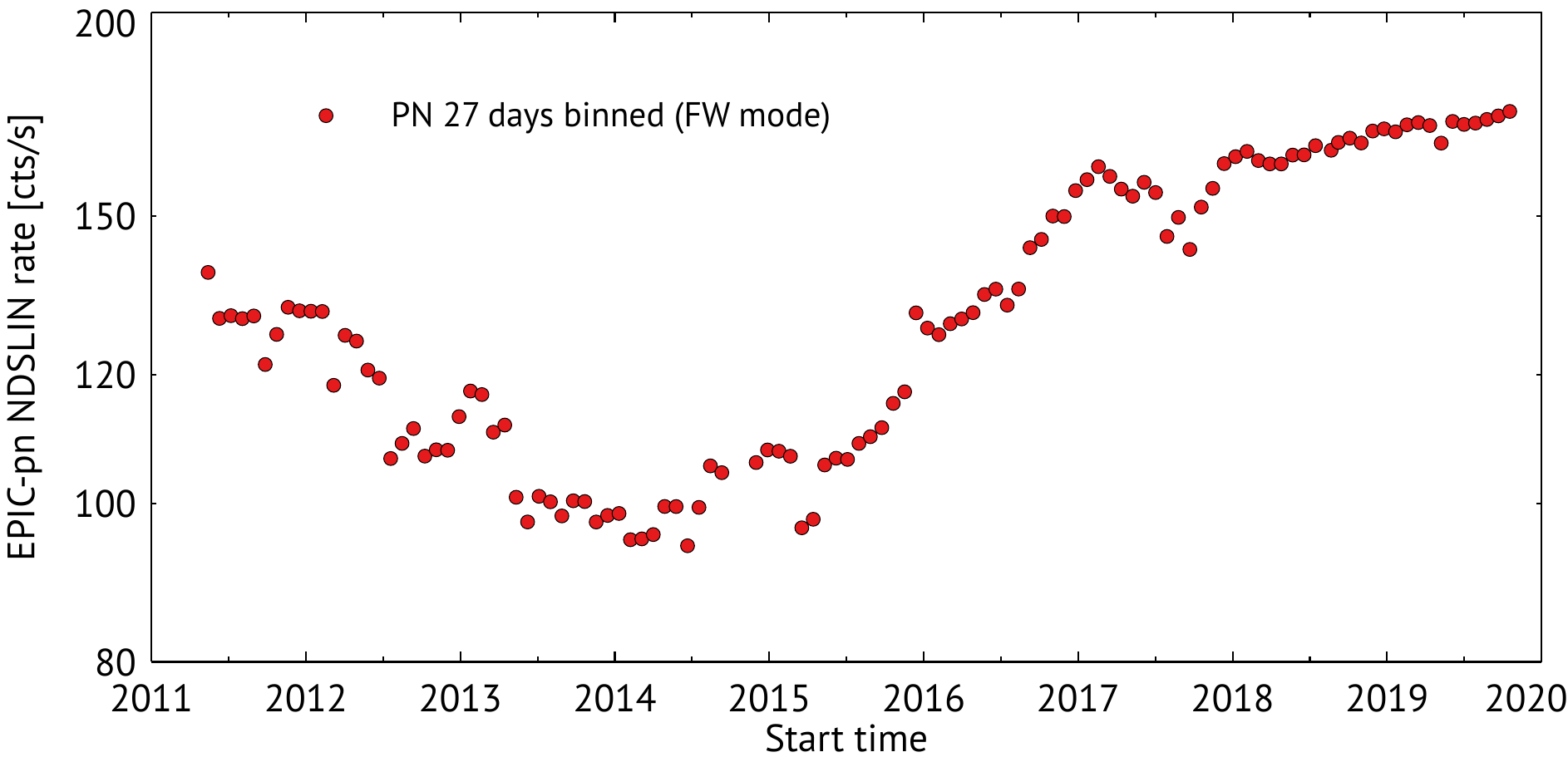}\\
    \end{tabular}
\caption{AMS proton flux ({\it top}),
ACIS high-energy reject rates ({\it middle}),
and EPIC-pn NDSLIN rates ({\it bottom})
binned over 27-days Bartel bin, spanning
the years 2011--2020.
The 1-$\sigma$ 
uncertainties 
are plotted, 
but are too small to see.
}
\label{fig:27days_binned}
\end{figure*}

To facilitate comparison across 
all three datasets, 
we have first re-binned them into 27-day intervals,
commonly referred to as Bartel's rotation.
Within the data, outliers 
correspond to distinct solar 
storms that led to heightened particle 
rates. 
Our primary interest lies in
investigating quiescent variations, 
prompting us to remove the positive
outliers
during the binning process. 
{
To detect and remove the outliers, 
we implemented a 3-$\sigma$
clipping method for 
the ACIS and EPIC-pn reject rates.
Each 27-day bin was
carefully examined, 
and count rates 
exceeding 3-$\sigma$ from 
the median value of the 
respective bin were flagged
as anomalous and subsequently 
removed.
}
However, negative outliers were not
removed,
as they indicate extended declines in
particle background following 
intense solar events 
\citep{1937PhRv...51.1108F}. 
Figure \ref{fig:27days_binned}
illustrates the outcome 
of this 27-day binning for the AMS,
ACIS, and EPIC-pn datasets. 
Each panel incorporates 1$\sigma$
uncertainties, 
although they might not be
clearly visible due to their small
size (comprising both statistical and 
systematic errors for AMS, and 
solely statistical errors 
for ACIS and PN).
Despite differences in uncertainties
and being plotted on different
y-ranges,
all three datasets exhibit 
noticeable degrees of
similar 
features characteristic of
the solar cycle,
encompassing a significant
portion of Solar Cycle 23 and a
smaller fraction of Cycle 24.

\begin{figure*}
    \centering
    \begin{tabular}{cc}
         \includegraphics[width=0.58\textwidth]{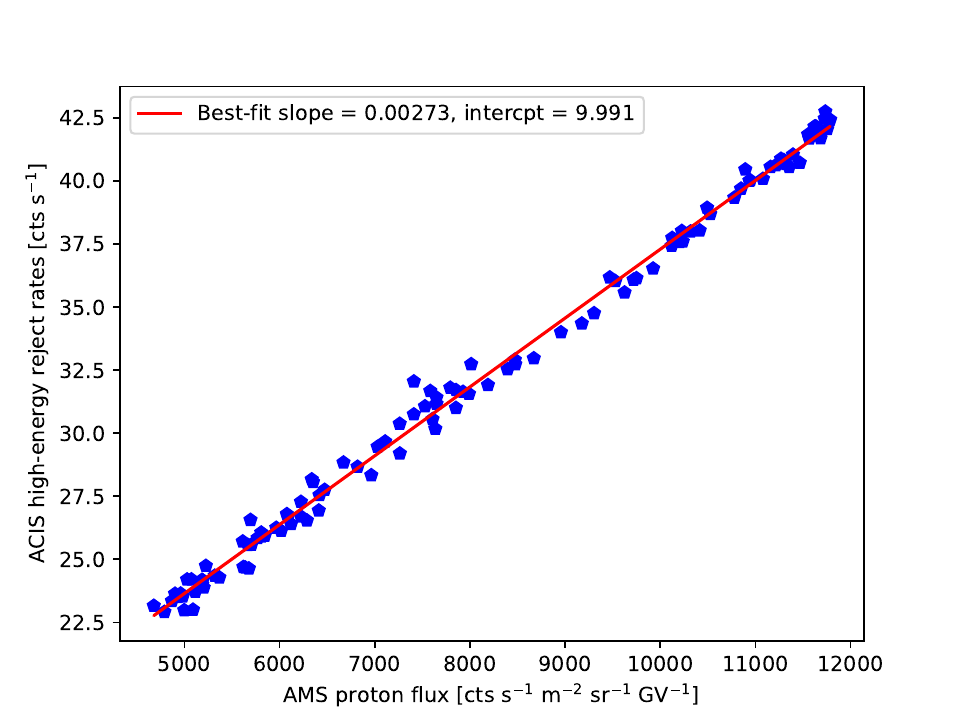} &
         \hspace{-40pt}
         \includegraphics[width=0.58\textwidth]{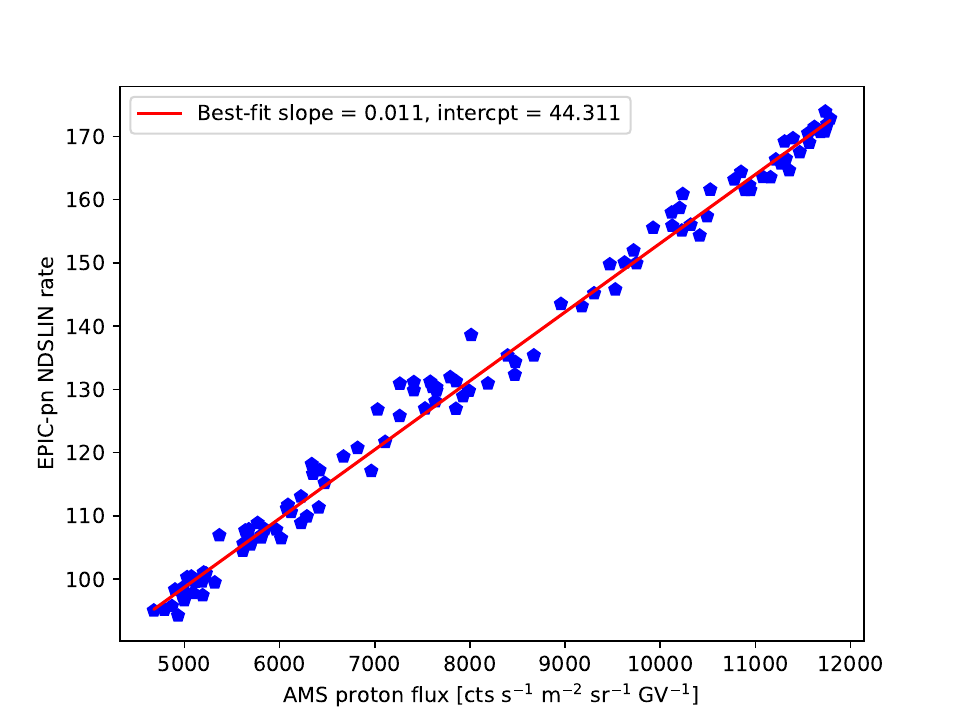}\\
    \end{tabular}
\caption{Correlation between
27-day binned datasets. {\it Left:}
ACIS high-energy reject rates as
a function of AMS proton flux.
{\it Right:} EPIC-pn NDSLIN rates
as a function fo AMS proton flux.
The best-fit line is shown in red. 
}
\label{fig:ams_vs_acis_vs_pn}
\end{figure*}

To further underscore this correlation, 
we have employed linear regression to 
model relationships between the
ACIS reject rates and 
AMS proton flux, as well as between
the EPIC-pn NDSLIN rates and AMS proton flux. 
Figure 
\ref{fig:ams_vs_acis_vs_pn}
provides a visualization 
of these datasets along with their 
corresponding best-fit linear
regression lines
{
($\chi^2$/dof = 43.6/111 for EPIC-pn
and $\chi^2$/dof = 24.8/111 for ACIS)
}. 
The computed best-fit slope for the 
AMS vs. ACIS fitting is 0.00273,
while for the AMS vs. EPIC-pn fitting, 
it is 0.011. 
Upon extrapolation to zero AMS proton 
flux, the linear fits yield 
{ non-zero intercepts}
ACIS count rates of approximately 10 
counts per second and PN count rates 
of roughly 44 counts per second. 
These non-zero intercepts are, 
in part, 
attributable to the consistent 
pedestal contributed by the hard 
X-ray background. 
A similar correlation
between ACIS reject
rates and AMS proton
flux was also found by
\citet{2022SPIE12181E..2EG}.
Notably, both linear fits exhibit high 
statistical significance. 
However, we observe minor scatter
around the best-fit lines, 
potentially stemming from actual
disparities between the two 
respective
datasets within each time bin.

\begin{figure*}
    \centering
    \begin{tabular}{c}
    \hspace{-50pt}
         \includegraphics[width=1.2\textwidth]{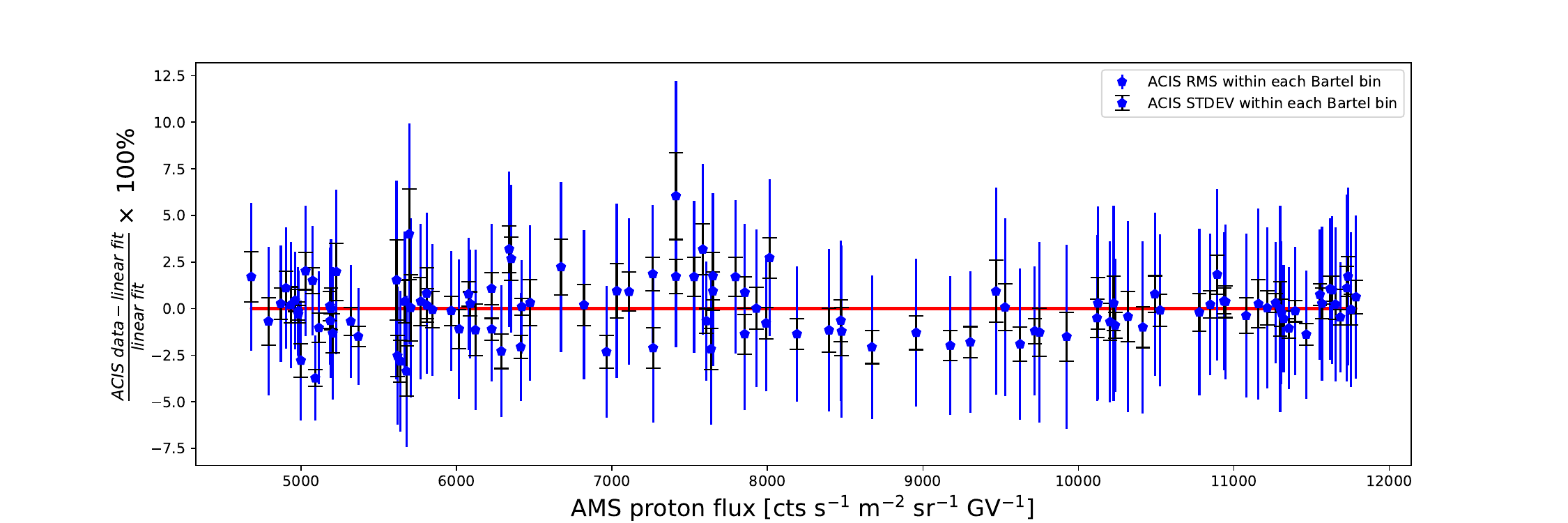}\\
         \hspace{-50pt}
         \includegraphics[width=1.2\textwidth]{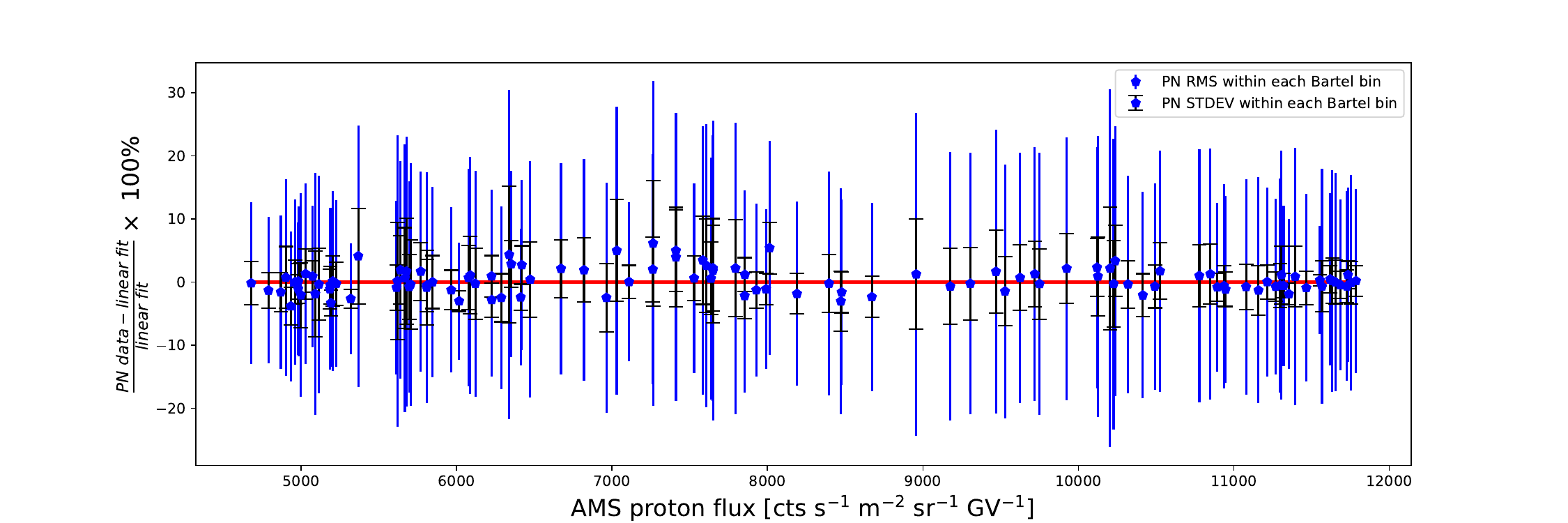}
    \end{tabular}
\caption{
{\it Top:} the difference between
the ACIS high-energy reject rates and the
best-fit line as a function of the 
AMS proton flux. Both datasets were
binned over 27-days.
The black uncertainties on each 
data points
represent the intrinsic scatter 
in ACIS data within each 27-day bin.
The blue uncertainties represent
RMS scatter of ACIS data within each 
27-day bin
{\it Bottom:} similar to the {\it Top:},
but between PN NDSLIN rates and AMS
proton flux.
}
\label{fig:rms}
\end{figure*}

To gain a deeper insight into 
the factors contributing to the 
scatter observed in the correlation,
we conducted an analysis of 
root mean square (RMS) deviations. 
Specifically, we computed the
RMS deviation of the ACIS and PN data 
within each 27-day time bin from their 
respective best-fit linear
regression lines.
Figure 
\ref{fig:rms}
illustrates the resultant
RMS deviations for each 27-day bin 
as a function of AMS proton flux.
While the overall correlation remains 
evident, the notable magnitude of
the RMS deviation underscores the 
existence of substantial variations
within the ACIS and PN datasets 
occurring on timescales shorter 
than 27 days. 
This observation 
strongly indicates that the 
fluctuations in GCR particle 
flux exhibit finer temporal 
variations than the 27-day 
interval under consideration.

\subsection{Daily binned data}\label{sec:daily_bin}
{ To examine the short term 
variation
in the particle background}, 
we re-binned the AMS proton flux,
ACIS reject rates, and 
PN NDSLIN rates into one-day intervals.
This re-binning strategy was chosen to
align with the minimal time bin of 
one day provided by 
publicly released AMS data, 
ensuring consistency and 
facilitating direct comparisons
across datasets. 
Additionally, for a 
comprehensive comparison 
of all three datasets, 
we re-normalized
the ACIS and PN rates
to match the AMS proton flux. 
This normalization was 
achieved using the
best-fit slopes and 
intercepts as outlined in 
Section \ref{sec:corr}.

\begin{figure*}
    \centering
    \begin{tabular}{c}
         \includegraphics[width=0.8\textwidth]{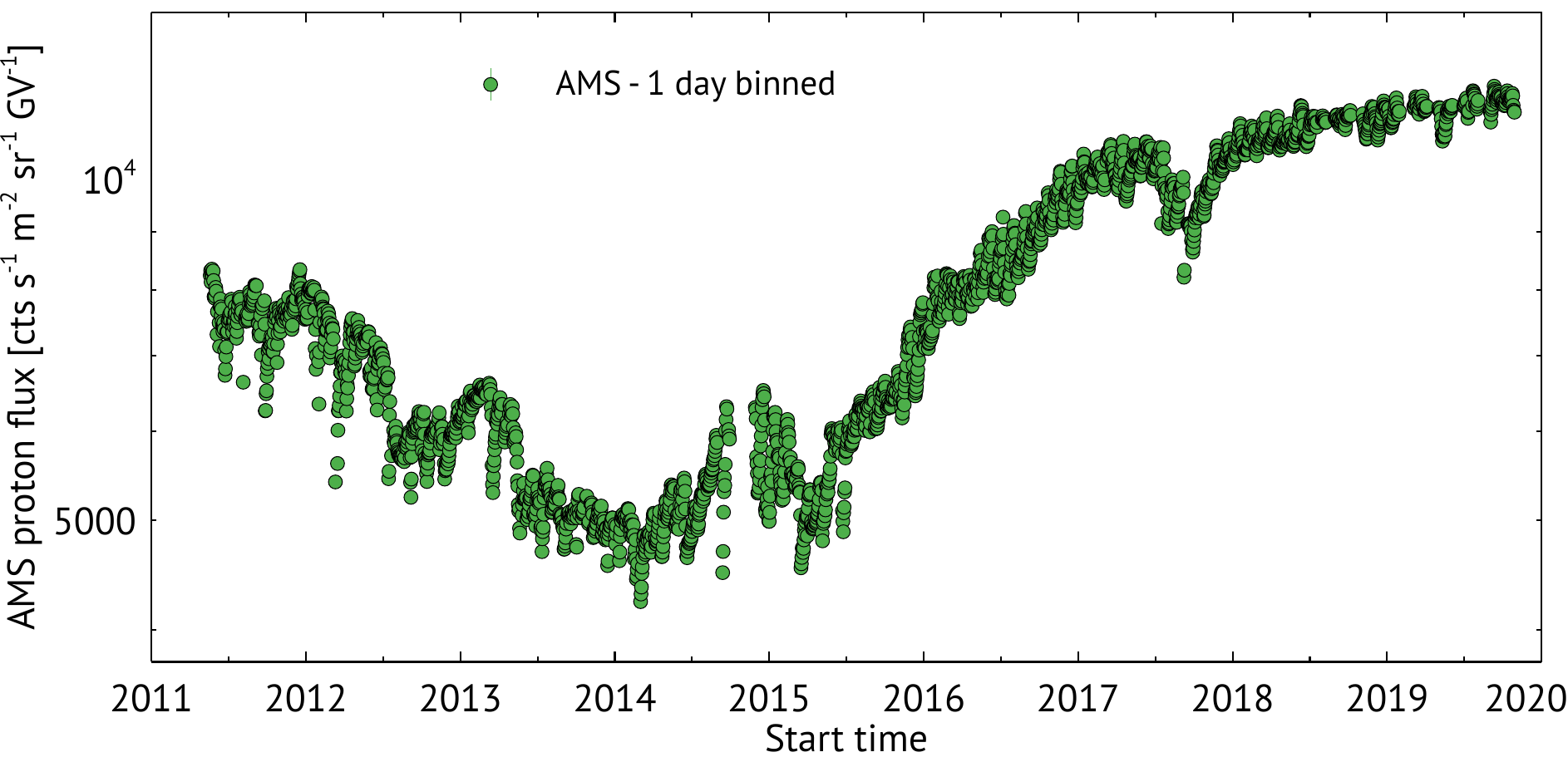} \\  \includegraphics[width=0.8\textwidth]{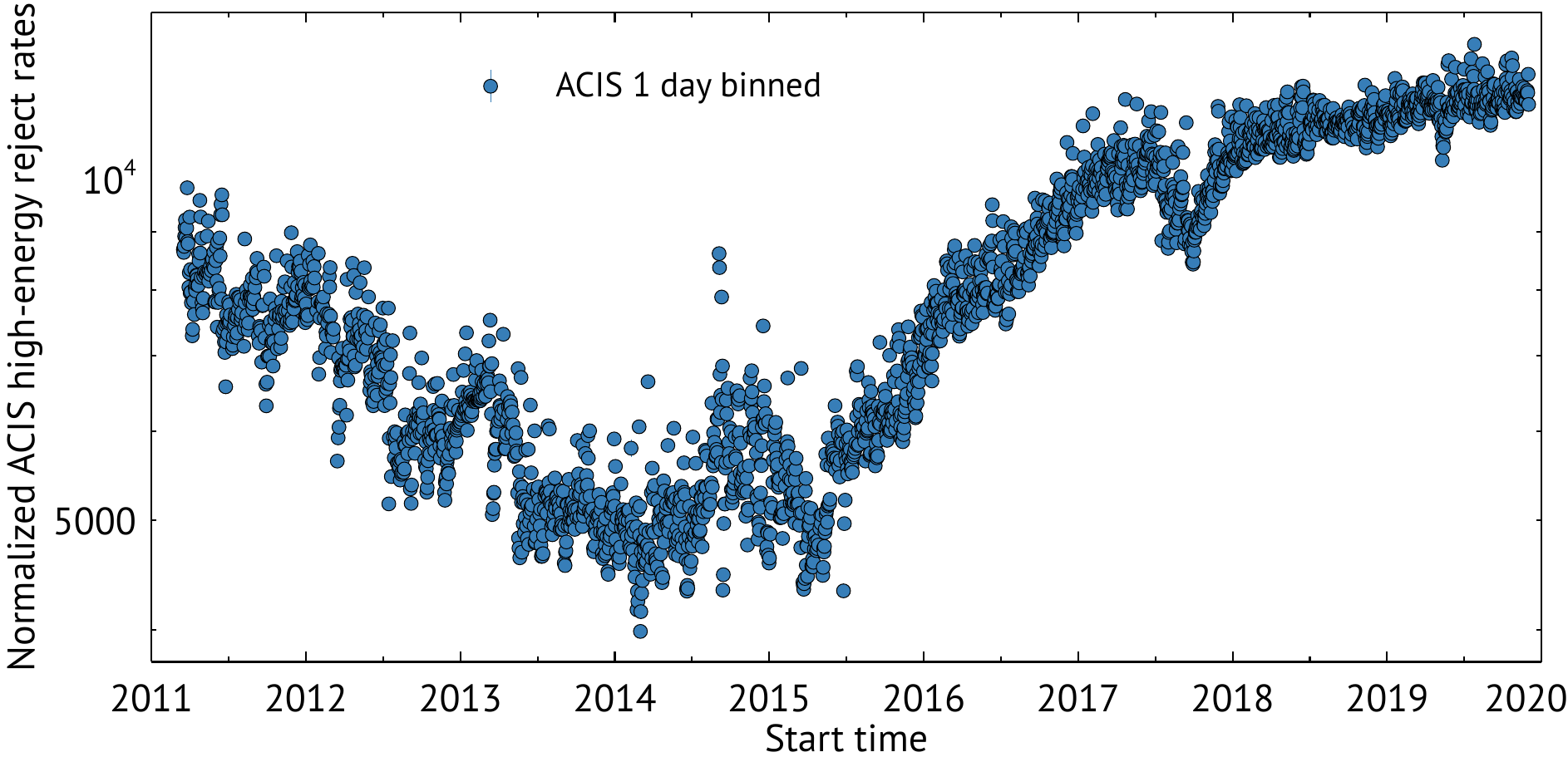}\\
         \includegraphics[width=0.8\textwidth]{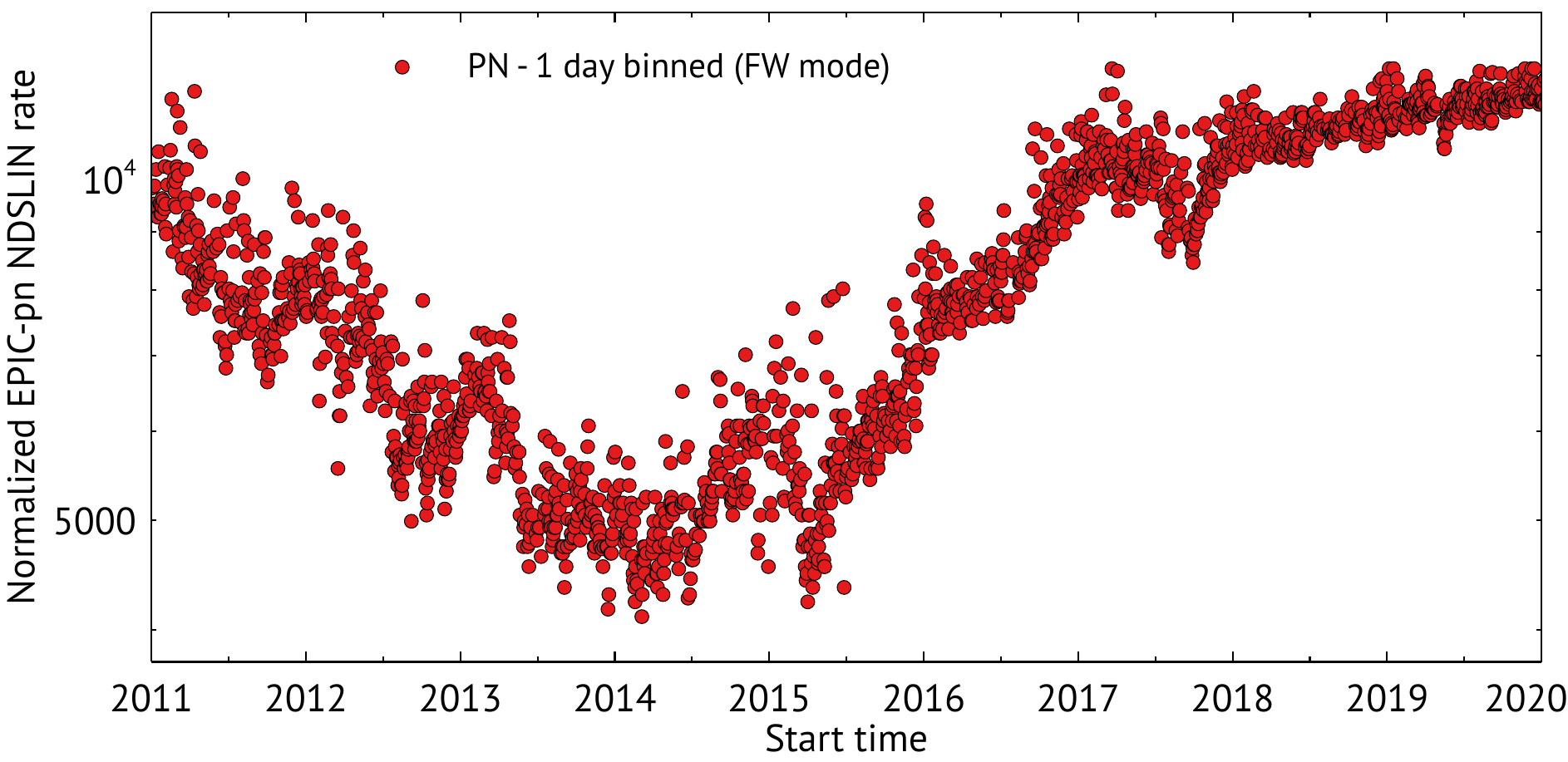}\\
    \end{tabular}
\caption{AMS proton flux, ({\it top})
ACIS high-energy reject rates ({\it middle}),
and EPIC-pn NDSLIN rates ({\it bottom})
binned over one-day, spanning
the years 2011--2020.
ACIS reject rates and EPIC-pn
NDSLIN rates were re-normalized
to match AMS proton flux
based on best-fit parameters
reported in Section 
\ref{sec:corr}.
The 1-$\sigma$ 
uncertainties 
are plotted, 
but are too small to see.}
\label{fig:1day_binned}
\end{figure*}

Figure \ref{fig:1day_binned}
visually presents the 
outcomes of this daily binning 
process for the AMS, ACIS, and 
PN datasets, encompassing the
timeframe from 2011 to 2020.
Each panel is accompanied by 
1$\sigma$ uncertainties, 
although their visual clarity
may be hindered by their relatively 
small size. 
For AMS data, these uncertainties 
encompass both statistical and 
systematic errors, whereas for 
ACIS and PN data, they solely 
encompass statistical errors. 
In handling outliers within
individual datasets, we adopt a 
methodology similar to that 
elaborated upon in the previous
section.

\begin{figure*}
    \centering
    \begin{tabular}{c}
\includegraphics[width=0.85\textwidth]{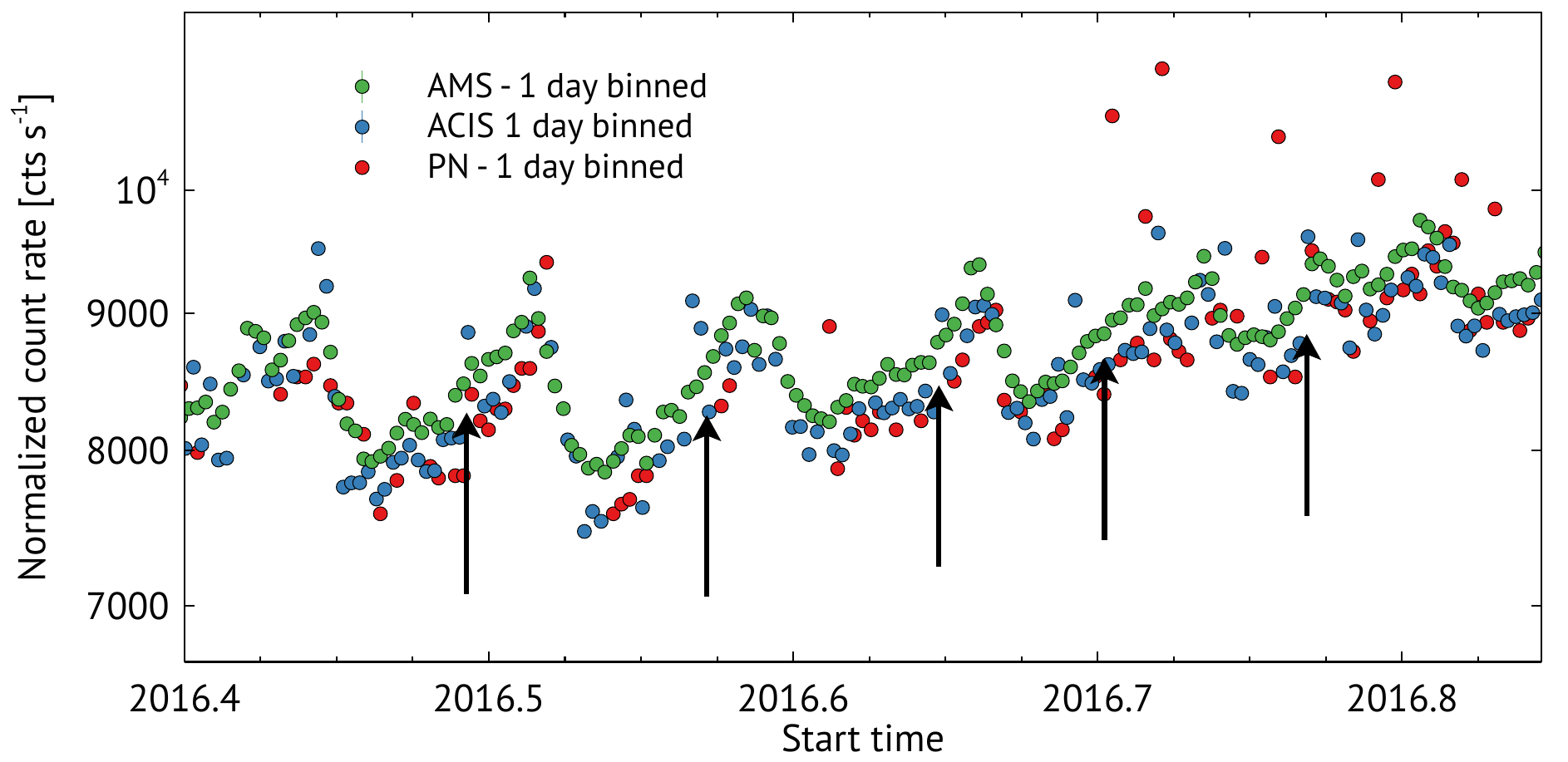}\\
\end{tabular}
\begin{tabular}{c}
\includegraphics[width=0.6\textwidth]{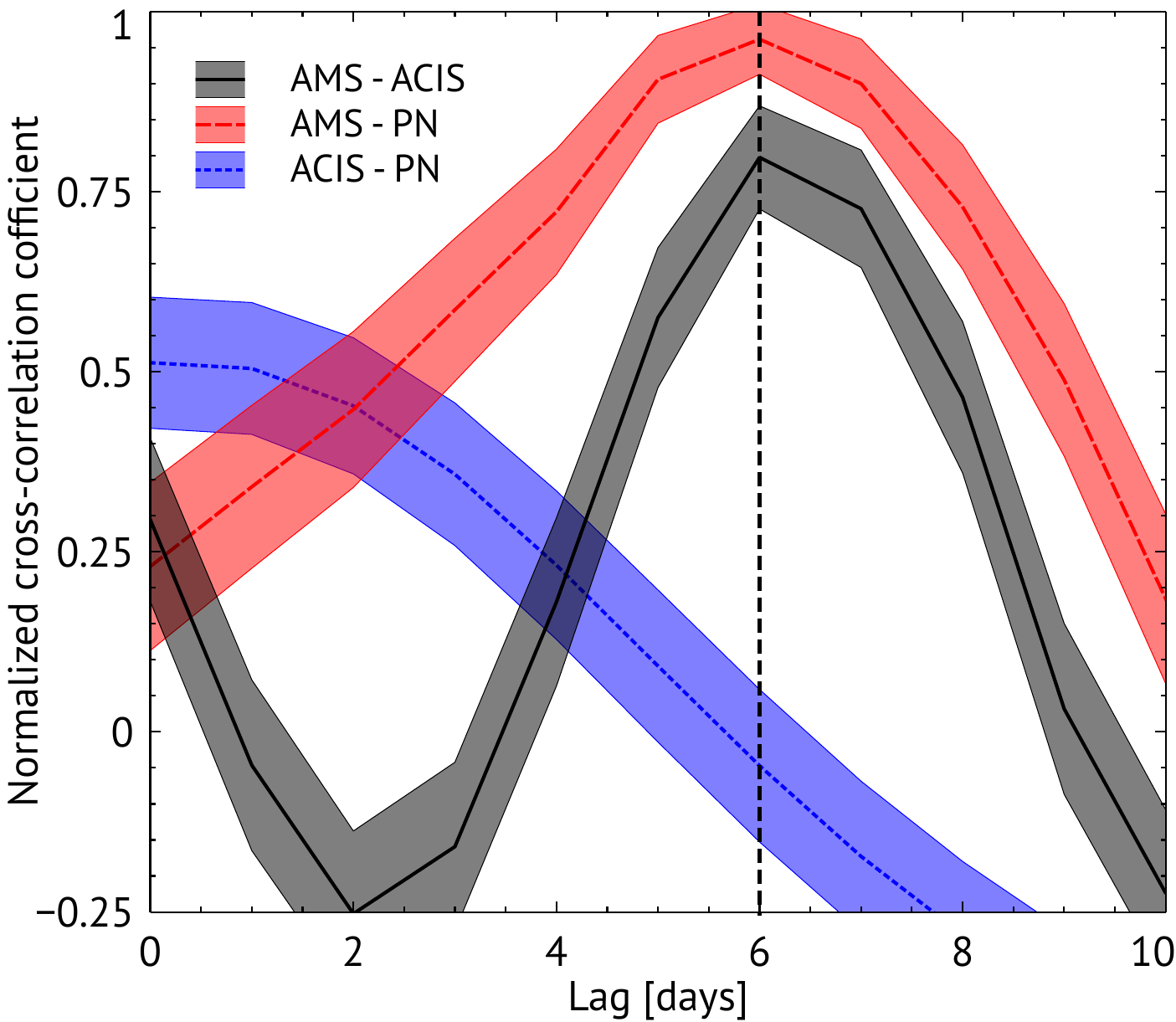}\\
\end{tabular}
\caption{
{\it Top:} one-day binned 
AMS proton flux, normalized ACIS
high-energy reject rates,
and normalized PN NDSLIN rates spanning
the years 2016--2017. 
The time lag of $\sim$ 6-days 
between AMS and ACIS/PN is marked
by black arrows.
{\it Bottom:} 
{ presents the cross-correlations of
AMS--ACIS, AMS--EPIC-pn,
and ACIS--EPIC-pn reject rates 
with 1$\sigma$ uncertainties,
respectively, in the similar
time-frame as top panel. 
X-axis shows
the time lag between two 
respective data-sets.
The 6-days lag has been
marked with vertical dashed
line.
}
}
\label{fig:1day_binned_2016_2017}
\end{figure*}

As expected from the AMS
data, the ACIS and pn data
exhibit statistically significant
variations on one-day time scales.
This difference is highlighted 
in Figure \ref{fig:1day_binned},
where the ACIS and 
PN datasets exhibit greater 
dispersion than the AMS dataset. 
This discrepancy can be attributed, 
in part, to the inherent 
characteristics of the respective 
instruments. 
AMS functions as a precision
GCR particle detector, 
while ACIS and PN tally the count
of events surpassing a specific 
energy threshold, acting as a 
proxy for the direct detection of
GCR particles. 
Consequently, a single GCR particle
can trigger multiple events 
(ranging from 1 to around 100)
in ACIS and PN, leading 
to intrinsic scattering in
these datasets 
\citep{2022JATIS...8a8001M}. 
Despite this inherent scattering, 
all three datasets illustrate 
notably similar temporal variations.
As depicted in Figure 
\ref{fig:1day_binned_2016_2017},
we present the temporal variation of
AMS proton flux, ACIS  reject 
rates, and PN NDSLIN rates spanning the 
years 2016 to 2017.
Evidently, all three datasets exhibit
a consistent sawtooth pattern 
characterized by comparable amplitudes 
and frequencies.
{ This observation clearly 
demonstrates that
day-to-day variations in cosmic ray flux
produce corresponding variation in the
instrumental background.}

\subsection{Periodicity in datasets}
The propagation of GCR 
particles from
the interstellar medium through the 
heliosphere 
is substantially influenced 
by the large-scale
solar wind 
flow and the turbulent 
interplanetary magnetic 
field (IMF), 
collectively referred to as 
``solar modulation''. 
Parker's transport equation 
\citep{1965SSRv....4..666P}
includes four 
principal modulation mechanisms, 
encompassing gradient and 
curvature drifts, diffusion
within the irregular IMF, 
radial convection in the 
expanding solar wind, and
adiabatic deceleration 
(or adiabatic energy loss) 
\citep[e.g.,][]{2000GeoRL..27.1823S,2016Ap&SS.361...44I,2021A&A...646A.128M}. 
These modulation processes 
intricately shape the 
temporal fluctuations evident 
in GCR particle fluxes, 
{ and
thus in principle play a role in 
modulating instrumental backgrounds.}

To investigate temporal 
variations across the broadest range of 
timescales, our search focus on the 1-day  
binned datasets. 
As illustrated in Figure 
\ref{fig:1day_binned}, 
the global GCR particle flux 
exhibits a minimum between 
2014 and 2015, with peaks 
before 2011 and around 2019. 
This long-term variability 
can be attributed to the 
modulation of GCR
particle fluxes, 
driven by the magnetic 
polarity reversal of the Sun's 
orientation 
\citep{2021ApJS..254...37F}.
The 22-year 
helio-magnetic cycle, 
known as the Hale cycle,
leads to a distinctive 
cosmic-ray intensity profile 
characterized by
alternating sharp and 
flat-topped features
\citep{2010GeoRL..3718101M}, 
as seen in Figure 
\ref{fig:1day_binned}. 
This periodic pattern has been 
extensively examined in the 
literature since 
the early 1960s, as
exemplified by studies such 
as \citet{2010GeoRL..3718101M} 
and references therein.

\begin{figure*}
    \centering
    \begin{tabular}{cc}
    \hspace{-15pt}
    \includegraphics[width=0.55\textwidth]{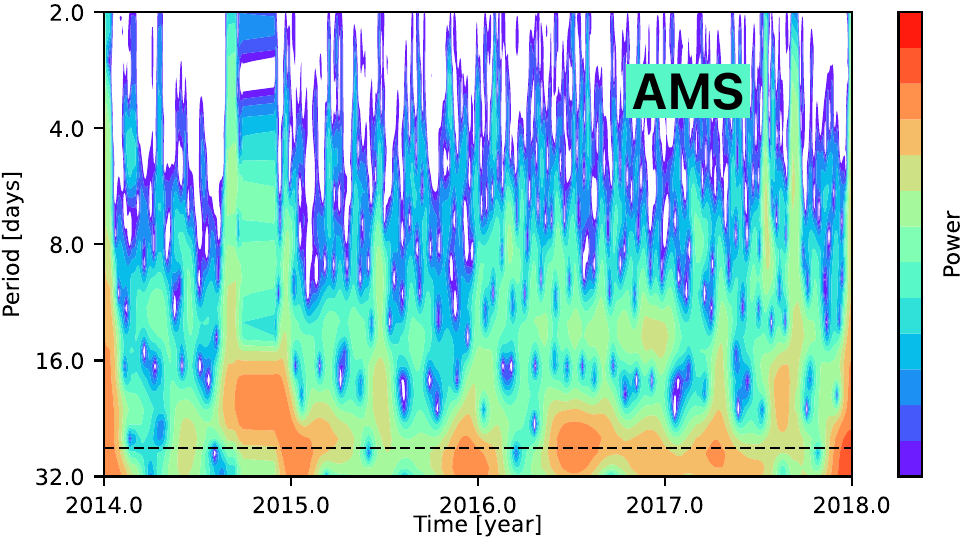} &
    \hspace{-40pt}
    \includegraphics[width=0.55\textwidth]{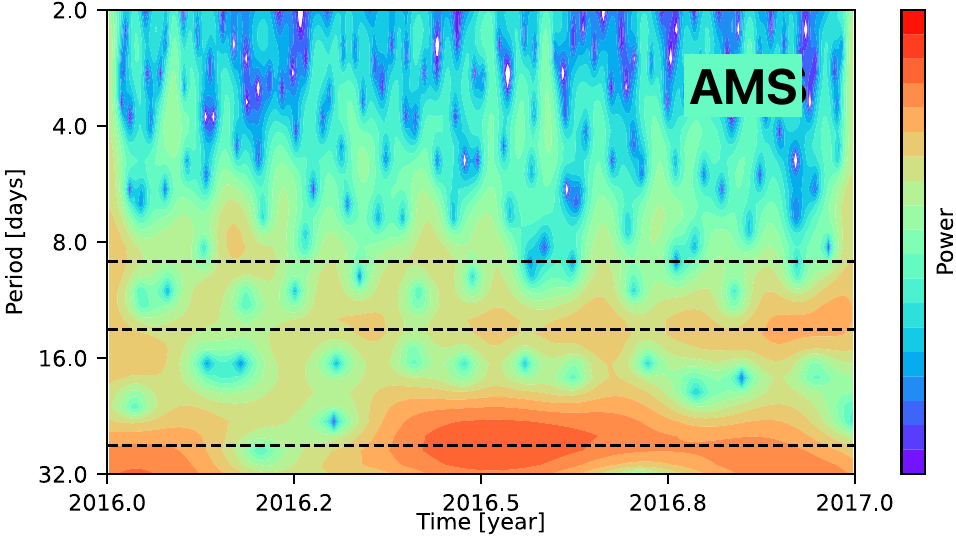} \\
     \hspace{-15pt}
     \includegraphics[width=0.55\textwidth]{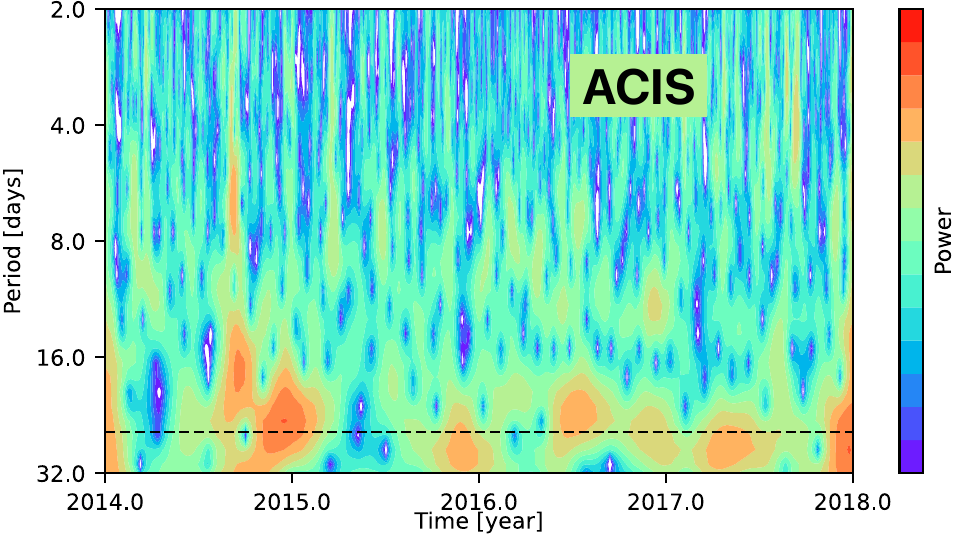} &
    \hspace{-40pt}
    \includegraphics[width=0.55\textwidth]{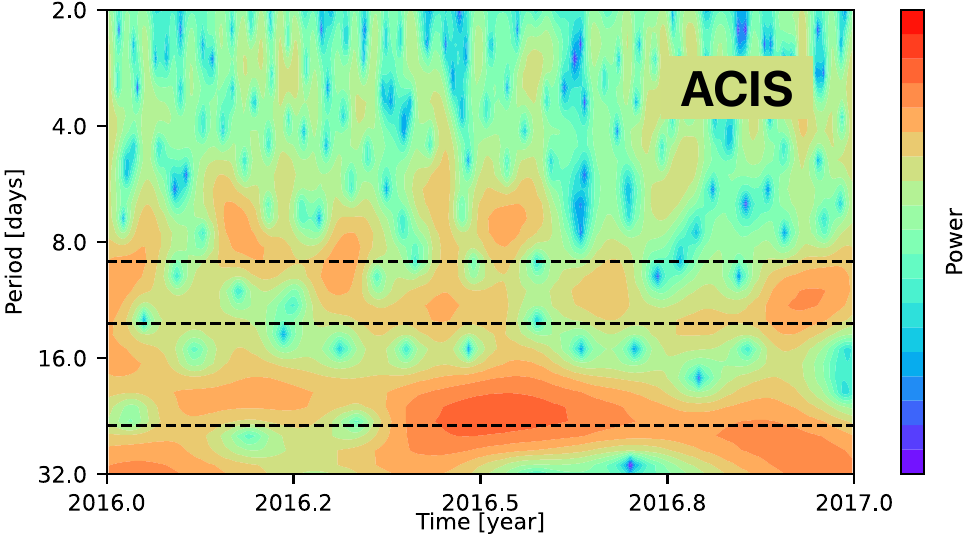} \\
    \hspace{-10pt}
    \includegraphics[width=0.56\textwidth]{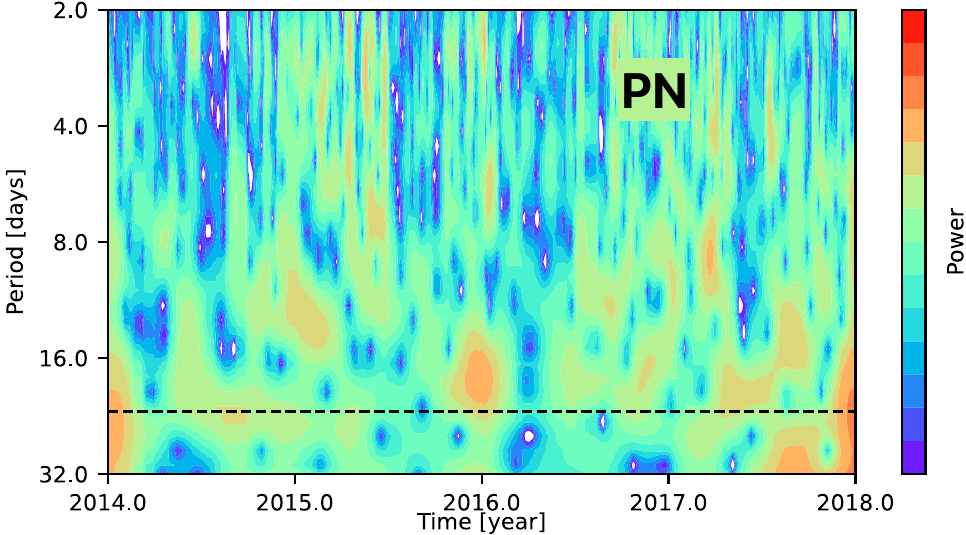} &
    \hspace{-40pt}
    \includegraphics[width=0.55\textwidth]{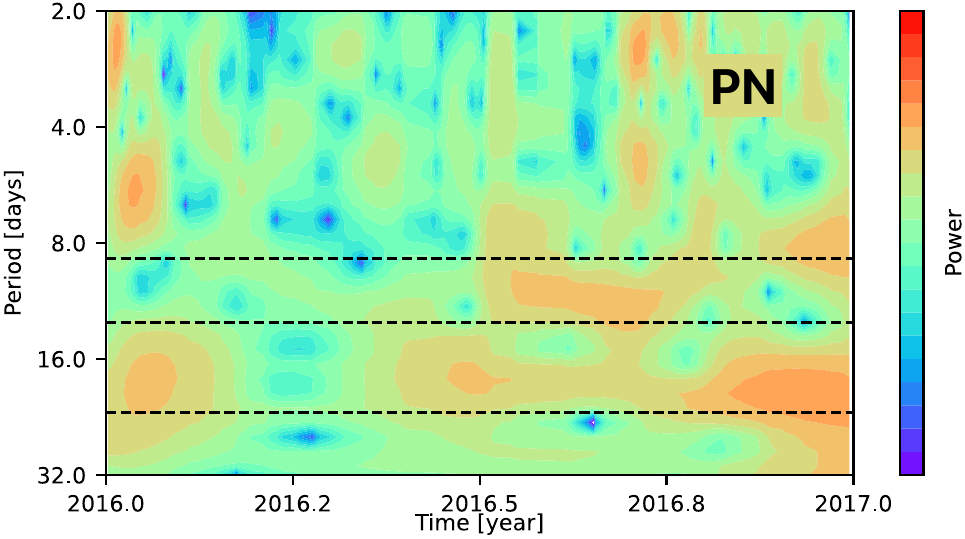} \\
    \end{tabular}
\caption{{\it Left panels (top to bottom):}
the wavelet time-frequency power
spectrum of one-day binned AMS proton
fluxes, ACIS high energy reject rates,
and PN NDSLIN rates, respectively,
spanning in the years
2014--2018.
The colorbar indicates the normalized power.
The horizontal
dashed lines indicate the 27 days,
25 days and 23 days periods
seen in AMS, ACIS, 
and PN datasets, respectively.
{\it Right panels:} similar to the
left panels, but zoomed in the year span
2016--2017. 
The horizontal
dashed lines indicate the 13.5 days,
9 days periods.
}
\label{fig:wavelet}
\end{figure*}

Previous studies 
employing daily-binned data 
have shown short-term
(ranging from months to even 
days) variability in AMS 
proton fluxes spanning the 
years 2014 to 2018 
\citep{2021PhRvL.127A1102A}, 
as partially depicted
in Figure 
\ref{fig:1day_binned_2016_2017}. 
This variability is 
particularly pronounced for
lower energies, specifically 
those less than approximately
10 GeV, 
where the heliosphere's 
magnetic shielding effect
becomes crucial. 
To probe
similar fluctuations 
within the ACIS and PN 
reject rates
over shorter timescales, 
we have employed the 
wavelet time-frequency 
technique, 
as described in 
\citet{1998BAMS...79...61T} and
\citet{2004NPGeo..11..561G}. 
A similar approach was
also employed 
by \citet{2021PhRvL.127A1102A}
while investigating 
periodicities in 
AMS proton flux.
Figure 
\ref{fig:wavelet}
visually 
presents the resultant 
wavelet power spectra for 
the daily-binned 
AMS proton flux, 
ACIS 
and PN reject rates, 
encompassing both the 
2014–2018 and 2016–2017 
year periods. 
All power spectra featured 
in subsequent figures are 
normalized { by the peak 
value} to effectively 
portray the strength of the 
underlying periodicities.

\citet{2021PhRvL.127A1102A} 
observed recurrent 
AMS proton flux variations 
with a period of $\sim$ 27 days 
at a significance above the 95\% 
confidence level from 2014 to 2018. 
Shorter periods 
of $\sim$13.5 days and 
$\sim$9 days were also
found only in 2016.
Both of these periodicities are 
clearly
visible in the top panels of Figure
\ref{fig:wavelet}. 
We have
identified a periodicity
of $\sim$ 25 days 
in ACIS high-energy reject
rates and a periodicity of 
$\sim$ 23 days 
in PN NDSLIN rates within
the 2014–2018 timeframe, 
as seen in the
middle-left and bottom-left
panels of Figure 
\ref{fig:wavelet}. 
Discrepancies in these periods 
compared to AMS observations 
are primarily attributed to 
intrinsic scattering present
in the ACIS and PN datasets
leading to choosing different peaks
in the datasets,
as described in Section 
\ref{sec:daily_bin}.

This recurring 27-day 
variation in GCR particle 
flux
was also observed on 
Earth by neutron monitors 
(NMs; \citealt{2013SoPh..286..593M}), 
in space by Advanced Composition 
Explorer 
(ACE; \citealt{2013SSRv..176..253L}), 
out of the ecliptic 
plane by Ulysses 
\citep{1995SSRv...72..403M}, 
and even in the remote
heliosphere  
by the Voyager spacecraft 
\citep{1999ICRC....7..512D}.
These observations point 
to the solar rotation as 
the dominant factor driving 
the 27-day variability in 
GCR particle flux. 
The uneven distribution of 
active regions on the Sun's 
surface lead to 
(heliographic) longitudinal 
asymmetry in electromagnetic 
conditions throughout one
solar rotation, thereby 
modulating the 27-day wave
observed in GCR particle fluxes 
\citep[e.g.,][]{2004SSRv..111..267R,2018LRSP...15....1R,2021A&A...646A.128M}.

Furthermore, we have also
identified a $\sim$ 
9-day and $\sim$ 
13.5-day variability in 
the ACIS 
and PN reject rates within
the 2016–2017 interval, similar
to the patterns observed in 
AMS data, as shown
in the right panels of 
Figure \ref{fig:wavelet}. 
These 13.5-day and 9-day
periodicities correspond 
to the higher harmonics of 
the $\sim$ 27-day GCR 
particle flux periodicity. 
Specifically, the 13.5-day 
periodicity represents the 
second harmonic, and 
the 9-day periodicity 
represents the third harmonic \citep{2004SSRv..111..267R,2011JGRA..116.4103S,2013SoPh..286..593M}. 
The quasi-period of $\sim$
13.5 days originates mainly due
to two distinct groups of 
active regions on the Sun's
surface, situated approximately
180 degrees apart in 
longitude 
\citep{2013SoPh..286..593M}. 
This second harmonic has
been meticulously explored
by 
\citet{1996JGR...10127077M}.

Furthermore, 
\citet{2011JGRA..116.4103S} 
showed that the effects of
corotating interaction 
regions, stemming from 
high-speed solar wind streams 
overtaking preceding slower 
solar wind, 
give rise to the 9-day
GCR modulations. 
Comparable variability of
$\sim$ 13.5 days and 9 days 
has also been observed in
Helios 1, Helios 2, and IMP-8 
experiments 
\citep{2004SSRv..111..267R}, 
Neutron Monitors 
\citep{2011JGRA..116.4103S}, 
as well as by
LISA Pathfinder 
\citep{2017JPhCS.840a2037G,2020ApJ...904...64G}.
These consistent findings
highlight the robust nature
of the observed periodicities
in the ACIS and PN reject rates.


\subsection{Time lag between datasets}
We have identified a time 
lag of $\sim$ 6 days 
between AMS proton flux and 
ACIS/EPIC-pn reject rates. 
This time lag exhibits 
greater prominence within 
the time span of 2016–2017, 
while it becomes less apparent 
prior to 2016 and 
subsequent to 2017. 
{
Figure 
\ref{fig:1day_binned_2016_2017}
top panel
provides a 
visual representation of this 
time-lag during the 
2016–-2017 period, 
indicated by black arrows.
To quantify
the time-lag, we estimate the 
cross-correlations between 
AMS proton flux and ACIS/EPIC-pn
reject rates.
The bottom panels of Figure 
\ref{fig:1day_binned_2016_2017} 
display the normalized 
cross-correlation coefficients 
and the corresponding time-lag
with 1$\sigma$ standard errors
(SE).
The standard errors were
estimated adopting,
$SE =\sqrt{\frac{1-r^2}{n}}$
\citep{Bonett2008},
where $r$ is cross-correlation
coefficient and $n$ is number
of data points.
A distinct time-lag of 6 days is 
evident in both cross-correlations, 
corroborating our observations 
from the top panel of
Figure 
\ref{fig:1day_binned_2016_2017}.
}
The underlying cause of this 
observed time lag 
remains uncertain. 


    
    To facilitate a 
    comprehensive comparison 
    of all three datasets,
    we have normalized the 
    ACIS and EPIC-pn reject rates
    to align with the AMS 
    proton flux
    using the best-fit 
    parameters obtained from
    the linear fit 
    presented in Figure 
    \ref{fig:ams_vs_acis_vs_pn}. 
    This normalization 
    process may introduce 
    slight short-term
    scaling discrepancies 
    in count rates. 
    This arises due to 
    the fact that the linear
    fitting was executed on 
    27-day binned data, 
    leading to a visual 
    impression of a time lag.
    It is important to 
    emphasize that  this
    lag is specifically 
    evident on the ascending 
    portions of the curve, 
    as depicted in Figure 
    \ref{fig:1day_binned_2016_2017}, 
    and is most noticeable in 
    the second half of 2016. 
    Notably, 
    within a similar time span,
    ACIS and EPIC-pn reject
    rates align with each
    other without any 
    noticeable lag, 
    despite being measured
    by different instruments
    and being scaled independently. 
    { We show in the
bottom panel of
Figure 
\ref{fig:1day_binned_2016_2017} that 
ACIS and EPIC-pn reject
    rates cross-correlates
best with no time lag.
}
    These observations suggest
    that the observed time 
    lag may not solely be 
    attributed to the 
    scaling factor.

\section{Discussion}
A fundamental objective of 
forthcoming X-ray observatories 
is the mapping the cosmic web 
through
deep exposures of faint 
diffuse
sources. 
Achieving this science goal
hinges on maintaining low and
well-determined
background levels and a 
comprehensive understanding of
the residual unrejected background. 
Notably, the predominant 
contributor to the particle
background 
at energy levels surpassing 
1-2 keV is attributed 
to GCR protons.
As described in the 
preceding section, the 
fluxes and spectrum of GCR 
protons exhibit modulation 
patterns related to both 
the 11-year solar cycle and 
shorter temporal scales 
characterized by spans of months, 
weeks, or even days due to solar 
activity.
The principal 
objective of this paper
centers on the understanding
this variabilty
in various locations in space,
notably those situated at
L1/L2, which
may prove crucial to 
minimizing uncertainties 
in particle-background 
of ESA’s NewAthena X-ray Observatory
and other missions with 
large collecting area,
like Lynx 
\citep{2019JATIS...5b1001G},
Advanced X-ray Imaging 
Satellite (AXIS; 
\citealt{2019BAAS...51g.107M}), 
and
Line Emission Mapper (LEM; 
\citealt{2022arXiv221109827K}).


Particle background data used in
analyzing faint
diffuse X-ray sources is often
derived
from the
extended observations 
($>$ 100 ks) 
of relatively blank 
part of the sky and  
dark Moon
\citep{2014A&A...566A..25B}.
However, a common challenge 
arises from the temporal 
disparity between the
blank-sky observation date 
and the actual observation of
the science target. 
The underlying assumption is
that the particle background 
remains stable at the
high elliptical orbit at 
any subsequent time post 
the blank-sky observation. 
We have shown that
the substantial variability of the 
GCR particle flux, 
on time scale as short as 1 day, 
produces corresponding variations 
in the background observed in 
the X-ray imaging instruments
(as seen in Figure \ref{fig:rms}).
An important question is, 
therefore,
the reliability of a
background measurement 
obtained at a 
specific time `$t$'
for utilization with
a science observation
at a later time. 
To answer this question, we 
have estimated the 
auto-correlation
of AMS proton flux, 
ACIS and EPIC-pn
reject rates starting from 
an
arbitrarily selected 
date. 
The auto-correlation function offers
a relative measure of the correlation 
between data at time t = 0 and 
any subsequent time.

\begin{figure}
    \centering
    \includegraphics[width=0.5\textwidth]{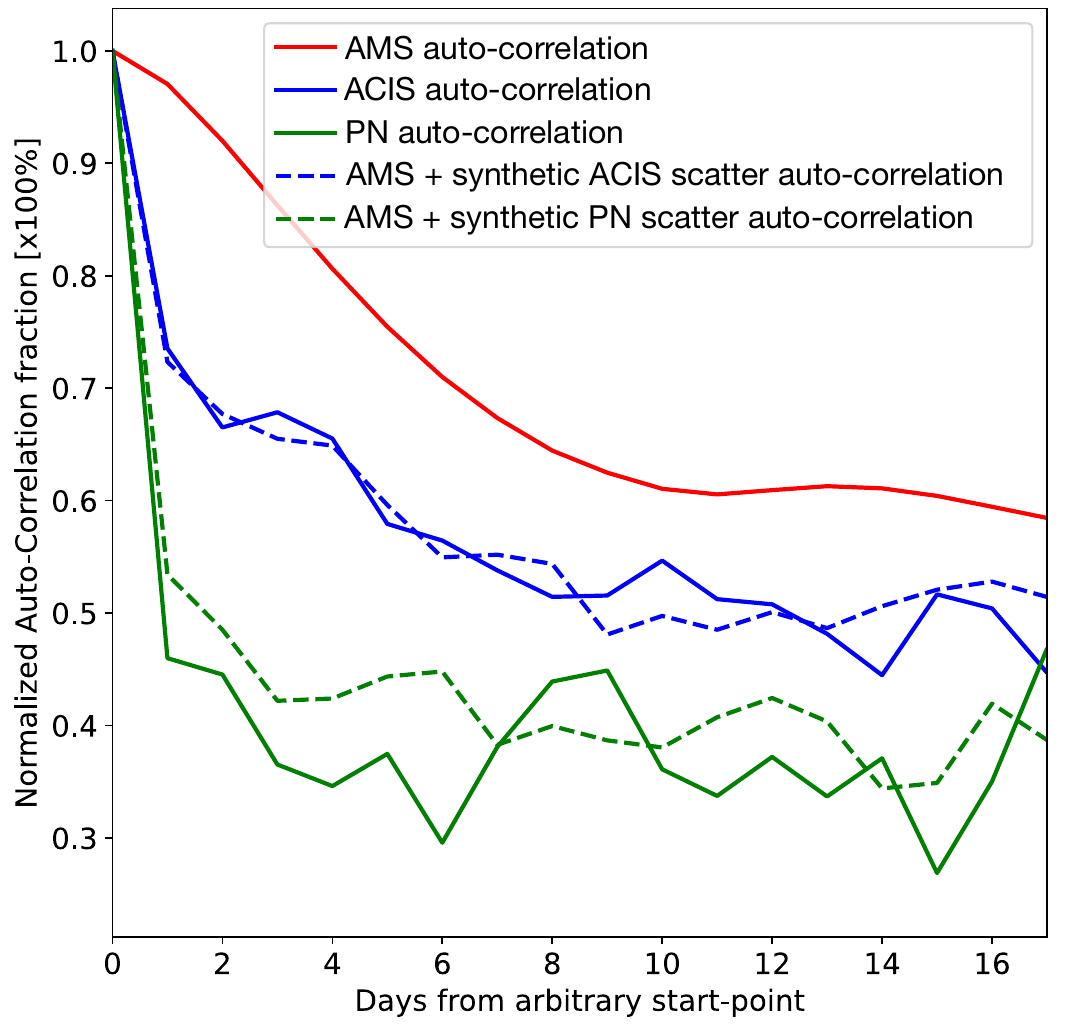}
\caption{
Auto-correlation functions of
AMS proton flux, ACIS high-energy
reject rates, and PN NDSLIN rates,
estimated using one-day binned data,
starting from an arbitrary date of
observation. The y-axis shows
the correlation percentage. 
{ Dashed blue and green curves show the
auto-correlation of 
the AMS proton flux after adding 
the synthetic scatter of ACIS and 
PN reject
rates, respectively.}
}
\label{fig:auto}
\end{figure}

For each of the datasets, 
we computed the 
auto-correlation between 
measurements at the 
initial day (t=0) and 
subsequent days. 
Figure 
\ref{fig:auto}
presents the 
resulting auto-correlation 
functions for all three datasets. 
Notably, the AMS dataset
maintains a significant
correlation percentage 
exceeding 70\% even after 
6 days from its initial 
measurement at t=0. 
However, the auto-correlation 
percentages for the ACIS and 
EPIC-pn reject rates exhibit 
a rapid decline compared to AMS. 
Specifically, for the 
ACIS dataset, the
auto-correlation drops
below 60\% by the 5th day, 
while for the EPIC-pn dataset, 
it falls below 50\% on
the subsequent day 
following the start-point. 

{ We investigated whether
the inherent fluctuations in the
ACIS and EPIC-pn reject rates are 
the main cause of the quicker 
decay observed in their 
auto-correlations compared to the AMS.
To explore this, 
we introduced random Gaussian noise
mimicking scatter in the ACIS and EPIC-pn
reject rates
to the AMS proton flux and 
computed the auto-correlation 
within the same time-frame. 
The auto-correlation of the AMS 
proton flux with added noise, 
illustrated by the dashed blue and green
curves in 
Figure \ref{fig:auto}, 
exhibited a similarly rapid decline 
as observed in the EPIC-pn and 
ACIS reject rates.
This finding confirms that the 
intrinsic variability in the 
EPIC-pn and ACIS reject rates is 
the primary contributor to the 
rapid decline in their 
auto-correlations. Moreover, it 
suggests that additional 
stochastic processes and sources
may contribute to the variability,
the mechanisms of which are 
not yet fully understood.
}

These observations pose a 
challenge to
the conventional practice of 
employing blank-sky observations 
to constrain particle background, 
especially considering its 
misalignment with the
actual date of scientific 
observations. 
{
This underscores the 
potential value of  of incorporating 
dedicated instrumentation to monitor background-generating
particles in X-ray telescopes, especially operating in orbits with little or no 
geomagnetic shielding.}
However, the AMS, serving 
as a precision GCR particle 
detector,
demonstrates the capacity to 
retain a higher correlation 
percentage, 
thereby holding potential 
for particle-background 
constraint for X-ray telescopes. 
In an upcoming study, 
we will show how the
AMS can effectively 
be utilized to predict
particle background for
telescopes situated in 
high elliptical orbits.

\section{Conclusion}
For the first time,
we have introduced 
daily-binned datasets 
encompassing high-energy
reject rates recorded by 
ACIS and EPIC-pn instruments,
using nearly all
archival science observations
spanning from 2011 to 2020.
These high-energy reject
rates serve as proxy 
measurements for GCR 
proton fluxes, which are 
precisely measured
by the AMS particle 
detector onboard the ISS.
We have conducted a 
comprehensive comparison 
of the variability observed 
in the ACIS and EPIC-pn reject
rates with that 
of AMS proton fluxes.  
Our findings are summarized below.
\begin{itemize}
   
    \item We compared all three 
    data sets binned to match 
    the 27 day binning in the
    AMS data release. 
    We find ACIS and EPIC-pn 
    reject
    rates strongly correlate 
    with 
    the AMS proton fluxes. 
    However, there is 
    substantial variation in the
    ACIS and EPIC-pn reject 
    rates 
    within each Bartel time 
    bin, 
    indicating variability 
    on shorter timescales. 
    This observation
    strongly indicates that the
    fluctuations in GCR particle flux, 
on shorter timescales, 
produces corresponding variations 
in the background observed in 
the X-ray imaging instruments.

    \item To understand the 
    temporal dynamics of the GCR particles,
    we rebinned ACIS and PN
    datasets into one-day interval, similar
    to AMS.
    Daily binned datasets exhibit 
    finer temporal structures in
    comparison to those binned over
    27-days intervals.
    {
    We note ACIS and EPIC-pn datasets
    display more intrinsic 
    scattering than AMS. 
    The excess variability in
    ACIS and EPIC-pn data 
    compared to AMS is of 
    unknown origin and may
    involve additional sources
    of instrumental background
    beyond those produced by
    GCR.} While AMS directly
    measure the GCR proton flux, ACIS
    and EPIC-pn counts the events
    surpassing a specific energy
    threshold. Despite this intrinsic
    scattering, all three
    datasets show remarkably similar
    temporal variations with similar
    amplitudes and frequencies.

    \item All three datasets show
    a overall minima between 2014--2015
    and peaks $<$ 2011 and 2019.
    The long-term variability is due 
    to the modulation of the GCR
    particle fluxes by the 
    11-year solar cycle. 
    We also find recurrent ACIS and
    EPIC-pn reject rates vary
    with a period of $\sim$ 25 days
    and $\sim$ 23 days, respectively,
    in the 2014--2018 year band.
    Similar variation in AMS
    proton flux was also observed
    by \citet{2021PhRvL.127A1102A}.
    The longitudinal asymmetry of
    the electro-magnetic field in
    the heliosphere, which arises 
    from the uneven distribution of 
    active regions on the Sun's 
    surface during a single solar 
    rotation, modulates the
    27-day period in GCR particle fluxes.
    Additionally, we have also
    detected $\sim$ 13.5 days and 
    $\sim$ 9 days period,
    also known as the second and
    third harmonics of 27-days period,
    in ACIS and
    EPIC-pn reject rates 
    between 2016--2017.
    
     \item A time lag of $\sim$ 6 days
     has been observed between AMS
     proton flux and ACIS/EPIC-pn 
     reject
     rates in the second half of 
     2016.
     This time lag is 
     predominantly absent 
     before 2016 and after 2017. 
     The underlying cause of
     this time lag remains 
     largely unidentified. 
     Nevertheless, the time 
     lag could potentially 
     stem from the 
     scaling relationship 
     between AMS and ACIS/EPIC-pn, 
     as well as the 
     reconstruction 
     procedures applied 
     to the AMS proton flux
     by the AMS team.

\end{itemize}

We thank anonymous referee for their helpful comments and suggestions.
We are grateful to the AMS team for their thorough documentation and the public availability of AMS data. This work was done as part of the Athena WFI Background Working Group, a consortium including MPE, INAF/IASF-Milano, IAAT, Open University, MIT, SAO, and Stanford. We gratefully acknowledge support from NASA cooperative agreement 80NSSC21M0046 and NASA contracts NAS 8-37716 and NAS 8-38252.

\bibliography{sample631}{}

\begin{thebibliography}{}
\expandafter\ifx\csname natexlab\endcsname\relax\def\natexlab#1{#1}\fi
\providecommand{\url}[1]{\href{#1}{#1}}
\providecommand{\dodoi}[1]{doi:~\href{http://doi.org/#1}{\nolinkurl{#1}}}
\providecommand{\doeprint}[1]{\href{http://ascl.net/#1}{\nolinkurl{http://ascl.net/#1}}}
\providecommand{\doarXiv}[1]{\href{https://arxiv.org/abs/#1}{\nolinkurl{https://arxiv.org/abs/#1}}}

\bibitem[{{Aguilar} {et~al.}(2015){Aguilar}, {Aisa}, {Alpat}, {Alvino},
  {Ambrosi}, {Andeen}, {Arruda}, {Attig}, {Azzarello}, {Bachlechner}, \&
  et~al.}]{2015PhRvL.114q1103A}
{Aguilar}, M., {Aisa}, D., {Alpat}, B., {et~al.} 2015, \prl, 114, 171103,
  \dodoi{10.1103/PhysRevLett.114.171103}

\bibitem[{{Aguilar} {et~al.}(2018){Aguilar}, {Ali Cavasonza}, {Alpat},
  {Ambrosi}, {Arruda}, {Attig}, {Aupetit}, {Azzarello}, {Bachlechner}, {Barao},
  \& et~al.}]{2018PhRvL.121e1101A}
{Aguilar}, M., {Ali Cavasonza}, L., {Alpat}, B., {et~al.} 2018, \prl, 121,
  051101, \dodoi{10.1103/PhysRevLett.121.051101}

\bibitem[{{Aguilar} {et~al.}(2021){Aguilar}, {Cavasonza}, {Ambrosi}, {Arruda},
  {Attig}, {Barao}, {Barrin}, {Bartoloni}, {Ba{\c{s}}e{\v{g}}mez-du Pree},
  {Battiston}, \& et~al.}]{2021PhRvL.127A1102A}
{Aguilar}, M., {Cavasonza}, L.~A., {Ambrosi}, G., {et~al.} 2021, \prl, 127,
  271102, \dodoi{10.1103/PhysRevLett.127.271102}

\bibitem[{{Bartalucci} {et~al.}(2014){Bartalucci}, {Mazzotta}, {Bourdin}, \&
  {Vikhlinin}}]{2014A&A...566A..25B}
{Bartalucci}, I., {Mazzotta}, P., {Bourdin}, H., \& {Vikhlinin}, A. 2014, \aap,
  566, A25, \dodoi{10.1051/0004-6361/201423443}

\bibitem[{{Bazilevskaya} {et~al.}(2014){Bazilevskaya}, {Broomhall}, {Elsworth},
  \& {Nakariakov}}]{2014SSRv..186..359B}
{Bazilevskaya}, G., {Broomhall}, A.~M., {Elsworth}, Y., \& {Nakariakov}, V.~M.
  2014, \ssr, 186, 359, \dodoi{10.1007/s11214-014-0068-0}

\bibitem[{Bonett(2008)}]{Bonett2008}
Bonett, D.~G. 2008, Psychological Methods, 13, 173–181,
  \dodoi{10.1037/a0012868}

\bibitem[{{Boschini} {et~al.}(2017){Boschini}, {Della Torre}, {Gervasi},
  {Grandi}, {J{\'o}hannesson}, {Kachelriess}, {La Vacca}, {Masi}, {Moskalenko},
  {Orlando}, {Ostapchenko}, {Pensotti}, {Porter}, {Quadrani}, {Rancoita},
  {Rozza}, \& {Tacconi}}]{2017ApJ...840..115B}
{Boschini}, M.~J., {Della Torre}, S., {Gervasi}, M., {et~al.} 2017, \apj, 840,
  115, \dodoi{10.3847/1538-4357/aa6e4f}

\bibitem[{{Bulbul} {et~al.}(2020){Bulbul}, {Kraft}, {Nulsen}, {Freyberg},
  {Miller}, {Grant}, {Bautz}, {Burrows}, {Allen}, {Eraerds}, {Fioretti},
  {Gastaldello}, {Ghirardini}, {Hall}, {Meidinger}, {Molendi}, {Rau},
  {Wilkins}, \& {Wilms}}]{2020ApJ...891...13B}
{Bulbul}, E., {Kraft}, R., {Nulsen}, P., {et~al.} 2020, \apj, 891, 13,
  \dodoi{10.3847/1538-4357/ab698a}

\bibitem[{{Campana}(2022)}]{2022hxga.book...39C}
{Campana}, R. 2022, in Handbook of X-ray and Gamma-ray Astrophysics, 39,
  \dodoi{10.1007/978-981-16-4544-0_28-1}

\bibitem[{{Chowdhury} {et~al.}(2016){Chowdhury}, {Kudela}, \&
  {Moon}}]{2016SoPh..291..581C}
{Chowdhury}, P., {Kudela}, K., \& {Moon}, Y.~J. 2016, \solphys, 291, 581,
  \dodoi{10.1007/s11207-015-0832-7}

\bibitem[{{Decker}(1999)}]{1999ICRC....7..512D}
{Decker}, R. 1999, in International Cosmic Ray Conference, Vol.~7, 26th
  International Cosmic Ray Conference (ICRC26), Volume 7, 512

\bibitem[{{Forbush}(1937)}]{1937PhRv...51.1108F}
{Forbush}, S.~E. 1937, Physical Review, 51, 1108,
  \dodoi{10.1103/PhysRev.51.1108.3}

\bibitem[{{Fraser} {et~al.}(2014){Fraser}, {Read}, {Sembay}, {Carter}, \&
  {Schyns}}]{2014MNRAS.445.2146F}
{Fraser}, G.~W., {Read}, A.~M., {Sembay}, S., {Carter}, J.~A., \& {Schyns}, E.
  2014, \mnras, 445, 2146, \dodoi{10.1093/mnras/stu1865}

\bibitem[{{Freyberg} {et~al.}(2004){Freyberg}, {Briel}, {Dennerl}, {Haberl},
  {Hartner}, {Pfeffermann}, {Kendziorra}, {Kirsch}, \&
  {Lumb}}]{2004SPIE.5165..112F}
{Freyberg}, M.~J., {Briel}, U.~G., {Dennerl}, K., {et~al.} 2004, in Society of
  Photo-Optical Instrumentation Engineers (SPIE) Conference Series, Vol. 5165,
  X-Ray and Gamma-Ray Instrumentation for Astronomy XIII, ed. K.~A. {Flanagan}
  \& O.~H.~W. {Siegmund}, 112--122, \dodoi{10.1117/12.506277}

\bibitem[{{Fu} {et~al.}(2021){Fu}, {Zhang}, {Zhao}, \&
  {Li}}]{2021ApJS..254...37F}
{Fu}, S., {Zhang}, X., {Zhao}, L., \& {Li}, Y. 2021, \apjs, 254, 37,
  \dodoi{10.3847/1538-4365/abf936}

\bibitem[{{Gaskin} {et~al.}(2019){Gaskin}, {Swartz}, {Vikhlinin}, {{\"O}zel},
  {Gelmis}, {Arenberg}, {Bandler}, {Bautz}, {Civitani}, {Dominguez}, {Eckart},
  {Falcone}, {Figueroa-Feliciano}, {Freeman}, {G{\"u}nther}, {Havey},
  {Heilmann}, {Kilaru}, {Kraft}, {McCarley}, {McEntaffer}, {Pareschi},
  {Purcell}, {Reid}, {Schattenburg}, {Schwartz}, {Schwartz}, {Tananbaum},
  {Tremblay}, {Zhang}, \& {Zuhone}}]{2019JATIS...5b1001G}
{Gaskin}, J.~A., {Swartz}, D.~A., {Vikhlinin}, A., {et~al.} 2019, Journal of
  Astronomical Telescopes, Instruments, and Systems, 5, 021001,
  \dodoi{10.1117/1.JATIS.5.2.021001}

\bibitem[{{Gastaldello} {et~al.}(2017){Gastaldello}, {Ghizzardi}, {Marelli},
  {Salvetti}, {Molendi}, {De Luca}, {Moretti}, {Rossetti}, \&
  {Tiengo}}]{2017ExA....44..321G}
{Gastaldello}, F., {Ghizzardi}, S., {Marelli}, M., {et~al.} 2017, Experimental
  Astronomy, 44, 321, \dodoi{10.1007/s10686-017-9549-y}

\bibitem[{{Gastaldello} {et~al.}(2022){Gastaldello}, {Marelli}, {Molendi},
  {Bartalucci}, {K{\"u}hl}, {Grant}, {Ghizzardi}, {Rossetti}, {De Luca}, \&
  {Tiengo}}]{2022ApJ...928..168G}
{Gastaldello}, F., {Marelli}, M., {Molendi}, S., {et~al.} 2022, \apj, 928, 168,
  \dodoi{10.3847/1538-4357/ac5403}

\bibitem[{{Grant} {et~al.}(2002){Grant}, {Bautz}, \&
  {Virani}}]{2002ASPC..262..401G}
{Grant}, C.~E., {Bautz}, M.~W., \& {Virani}, S.~N. 2002, in Astronomical
  Society of the Pacific Conference Series, Vol. 262, The High Energy Universe
  at Sharp Focus: Chandra Science, ed. E.~M. {Schlegel} \& S.~D. {Vrtilek},
  401, \dodoi{10.48550/arXiv.astro-ph/0202086}

\bibitem[{{Grant} {et~al.}(2018){Grant}, {Miller}, {Bautz}, {Bulbul}, {Kraft},
  {Nulsen}, {Burrows}, \& {Allen}}]{2018SPIE10699E..4HG}
{Grant}, C.~E., {Miller}, E.~D., {Bautz}, M.~W., {et~al.} 2018, in Society of
  Photo-Optical Instrumentation Engineers (SPIE) Conference Series, Vol. 10699,
  Space Telescopes and Instrumentation 2018: Ultraviolet to Gamma Ray, ed.
  J.-W.~A. {den Herder}, S.~{Nikzad}, \& K.~{Nakazawa}, 106994H,
  \dodoi{10.1117/12.2313864}

\bibitem[{{Grant} {et~al.}(2022){Grant}, {Miller}, {Bautz}, {Foster}, {Kraft},
  {Allen}, \& {Burrows}}]{2022SPIE12181E..2EG}
{Grant}, C.~E., {Miller}, E.~D., {Bautz}, M.~W., {et~al.} 2022, in Society of
  Photo-Optical Instrumentation Engineers (SPIE) Conference Series, Vol. 12181,
  Space Telescopes and Instrumentation 2022: Ultraviolet to Gamma Ray, ed.
  J.-W.~A. {den Herder}, S.~{Nikzad}, \& K.~{Nakazawa}, 121812E,
  \dodoi{10.1117/12.2629520}

\bibitem[{{Grimani} {et~al.}(2020){Grimani}, {Cesarini}, {Fabi}, {Sabbatini},
  {Telloni}, \& {Villani}}]{2020ApJ...904...64G}
{Grimani}, C., {Cesarini}, A., {Fabi}, M., {et~al.} 2020, \apj, 904, 64,
  \dodoi{10.3847/1538-4357/abbb90}

\bibitem[{{Grimani} {et~al.}(2017){Grimani}, {LISA Pathfinder Collaboration},
  {Benella}, {Fabi}, {Finetti}, \& {Telloni}}]{2017JPhCS.840a2037G}
{Grimani}, C., {LISA Pathfinder Collaboration}, {Benella}, S., {et~al.} 2017,
  in Journal of Physics Conference Series, Vol. 840, Journal of Physics
  Conference Series, 012037, \dodoi{10.1088/1742-6596/840/1/012037}

\bibitem[{{Grinsted} {et~al.}(2004){Grinsted}, {Moore}, \&
  {Jevrejeva}}]{2004NPGeo..11..561G}
{Grinsted}, A., {Moore}, J.~C., \& {Jevrejeva}, S. 2004, Nonlinear Processes in
  Geophysics, 11, 561, \dodoi{10.5194/npg-11-561-2004}

\bibitem[{{Ihongo} \& {Wang}(2016)}]{2016Ap&SS.361...44I}
{Ihongo}, G.~D., \& {Wang}, C.~H.~T. 2016, \apss, 361, 44,
  \dodoi{10.1007/s10509-015-2628-5}

\bibitem[{{Kounine}(2012)}]{2012IJMPE..2130005K}
{Kounine}, A. 2012, International Journal of Modern Physics E, 21, 1230005,
  \dodoi{10.1142/S0218301312300056}

\bibitem[{{Kraft} {et~al.}(2022){Kraft}, {Markevitch}, {Kilbourne}, {Adams},
  {Akamatsu}, {Ayromlou}, {Bandler}, {Barbera}, {Bennett}, {Bhardwaj}, {Biffi},
  {Bodewits}, {Bogdan}, {Bonamente}, {Borgani}, {Branduardi-Raymont},
  {Bregman}, {Burchett}, {Cann}, {Carter}, {Chakraborty}, {Churazov}, {Crain},
  {Cumbee}, {Dave}, {DiPirro}, {Dolag}, {Bertrand Doriese}, {Drake}, {Dunn},
  {Eckart}, {Eckert}, {Ettori}, {Forman}, {Galeazzi}, {Gall}, {Gatuzz}, {Hell},
  {Hodges-Kluck}, {Jackman}, {Jahromi}, {Jennings}, {Jones}, {Kaaret},
  {Kavanagh}, {Kelley}, {Khabibullin}, {Kim}, {Koutroumpa}, {Kovacs}, {Kuntz},
  {Lau}, {Lee}, {Leutenegger}, {Lin}, {Lisse}, {Lo Cicero}, {Lovisari},
  {McCammon}, {McEntee}, {Mernier}, {Miller}, {Nagai}, {Negro}, {Nelson},
  {Ness}, {Nulsen}, {Ogorzalek}, {Oppenheimer}, {Oskinova}, {Patnaude},
  {Pfeifle}, {Pillepich}, {Plucinsky}, {Pooley}, {Porter}, {Randall}, {Rasia},
  {Raymond}, {Ruszkowski}, {Sakai}, {Sarkar}, {Sasaki}, {Sato},
  {Schellenberger}, {Schaye}, {Simionescu}, {Smith}, {Steiner}, {Stern}, {Su},
  {Sun}, {Tremblay}, {Truong}, {Tutt}, {Ursino}, {Veilleux}, {Vikhlinin},
  {Vladutescu-Zopp}, {Vogelsberger}, {Walker}, {Weaver}, {Weigt}, {Werk},
  {Werner}, {Wolk}, {Zhang}, {Zhang}, {Zhuravleva}, \&
  {ZuHone}}]{2022arXiv221109827K}
{Kraft}, R., {Markevitch}, M., {Kilbourne}, C., {et~al.} 2022, arXiv e-prints,
  arXiv:2211.09827, \dodoi{10.48550/arXiv.2211.09827}

\bibitem[{{Kudela} \& {Sabbah}(2016)}]{2016ScChE..59..547K}
{Kudela}, K., \& {Sabbah}, I. 2016, Science in China E: Technological Sciences,
  59, 547, \dodoi{10.1007/s11431-015-5924-y}

\bibitem[{{Kuntz} \& {Snowden}(2008)}]{2008A&A...478..575K}
{Kuntz}, K.~D., \& {Snowden}, S.~L. 2008, \aap, 478, 575,
  \dodoi{10.1051/0004-6361:20077912}

\bibitem[{{Leske} {et~al.}(2013){Leske}, {Cummings}, {Mewaldt}, \&
  {Stone}}]{2013SSRv..176..253L}
{Leske}, R.~A., {Cummings}, A.~C., {Mewaldt}, R.~A., \& {Stone}, E.~C. 2013,
  \ssr, 176, 253, \dodoi{10.1007/s11214-011-9772-1}

\bibitem[{{Lumb} {et~al.}(2002){Lumb}, {Warwick}, {Page}, \& {De
  Luca}}]{2002A&A...389...93L}
{Lumb}, D.~H., {Warwick}, R.~S., {Page}, M., \& {De Luca}, A. 2002, \aap, 389,
  93, \dodoi{10.1051/0004-6361:20020531}

\bibitem[{{Marelli} {et~al.}(2021){Marelli}, {Molendi}, {Rossetti},
  {Gastaldello}, {Salvetti}, {De Luca}, {Bartalucci}, {K{\"u}hl}, {Esposito},
  {Ghizzardi}, \& {Tiengo}}]{2021ApJ...908...37M}
{Marelli}, M., {Molendi}, S., {Rossetti}, M., {et~al.} 2021, \apj, 908, 37,
  \dodoi{10.3847/1538-4357/abcfbc}

\bibitem[{{McDonald} {et~al.}(2010){McDonald}, {Webber}, \&
  {Reames}}]{2010GeoRL..3718101M}
{McDonald}, F.~B., {Webber}, W.~R., \& {Reames}, D.~V. 2010, \grl, 37, L18101,
  \dodoi{10.1029/2010GL044218}

\bibitem[{{McKibben} {et~al.}(1995){McKibben}, {Simpson}, {Zhang}, {Bame}, \&
  {Balogh}}]{1995SSRv...72..403M}
{McKibben}, R.~B., {Simpson}, J.~A., {Zhang}, M., {Bame}, S., \& {Balogh}, A.
  1995, \ssr, 72, 403, \dodoi{10.1007/BF00768812}

\bibitem[{{Miller} {et~al.}(2022){Miller}, {Grant}, {Bautz}, {Molendi},
  {Kraft}, {Nulsen}, {Bulbul}, {Allen}, {Burrows}, {Eraerds}, {Fioretti},
  {Gastaldello}, {Hall}, {Hubbard}, {Keelan}, {Meidinger}, {Perinati}, {Rau},
  \& {Wilkins}}]{2022JATIS...8a8001M}
{Miller}, E.~D., {Grant}, C.~E., {Bautz}, M.~W., {et~al.} 2022, Journal of
  Astronomical Telescopes, Instruments, and Systems, 8, 018001,
  \dodoi{10.1117/1.JATIS.8.1.018001}

\bibitem[{{Modzelewska} \& {Alania}(2013)}]{2013SoPh..286..593M}
{Modzelewska}, R., \& {Alania}, M.~V. 2013, \solphys, 286, 593,
  \dodoi{10.1007/s11207-013-0261-4}

\bibitem[{{Modzelewska} \& {Gil}(2021)}]{2021A&A...646A.128M}
{Modzelewska}, R., \& {Gil}, A. 2021, \aap, 646, A128,
  \dodoi{10.1051/0004-6361/202039651}

\bibitem[{{Molendi}(2017)}]{2017ExA....44..263M}
{Molendi}, S. 2017, Experimental Astronomy, 44, 263,
  \dodoi{10.1007/s10686-017-9544-3}

\bibitem[{{Mursula} \& {Zieger}(1996)}]{1996JGR...10127077M}
{Mursula}, K., \& {Zieger}, B. 1996, \jgr, 101, 27077,
  \dodoi{10.1029/96JA02470}

\bibitem[{{Mushotzky} {et~al.}(2019){Mushotzky}, {Aird}, {Barger},
  {Cappelluti}, {Chartas}, {Corrales}, {Eufrasio}, {Fabian}, {Falcone},
  {Gallo}, {Gilli}, {Grant}, {Hardcastle}, {Hodges-Kluck}, {Kara}, {Koss},
  {Li}, {Lisse}, {Loewenstein}, {Markevitch}, {Meyer}, {Miller}, {Mulchaey},
  {Petre}, {Ptak}, {Reynolds}, {Russell}, {Safi-Harb}, {Smith}, {Snios},
  {Tombesi}, {Valencic}, {Walker}, {Williams}, {Winter}, {Yamaguchi}, {Zhang},
  {Arenberg}, {Brandt}, {Burrows}, {Georganopoulos}, {Miller}, {Norman}, \&
  {Rosati}}]{2019BAAS...51g.107M}
{Mushotzky}, R., {Aird}, J., {Barger}, A.~J., {et~al.} 2019, in Bulletin of the
  American Astronomical Society, Vol.~51, 107,
  \dodoi{10.48550/arXiv.1903.04083}

\bibitem[{{Nandra} {et~al.}(2013){Nandra}, {Barret}, {Barcons}, {Fabian}, {den
  Herder}, {Piro}, {Watson}, {Adami}, {Aird}, {Afonso}, \&
  et~al.}]{2013arXiv1306.2307N}
{Nandra}, K., {Barret}, D., {Barcons}, X., {et~al.} 2013, arXiv e-prints,
  arXiv:1306.2307, \dodoi{10.48550/arXiv.1306.2307}

\bibitem[{{Parker}(1965)}]{1965SSRv....4..666P}
{Parker}, E.~N. 1965, \ssr, 4, 666, \dodoi{10.1007/BF00216273}

\bibitem[{Poliszczuk {et~al.}(2023)Poliszczuk, Wilkins, Allen, Miller,
  Chattopadhyay, Bautz, Darve, Foster, Grant, Herrmann, Kraft, Morris, Orel,
  Sarkar, \& Schneider}]{10.1117/12.2677095}
Poliszczuk, A., Wilkins, D., Allen, S.~W., {et~al.} 2023, in Applications of
  Machine Learning 2023, ed. M.~E. Zelinski, T.~M. Taha, J.~Howe, \& B.~N.
  Narayanan, Vol. 12675, International Society for Optics and Photonics (SPIE),
  126750C, \dodoi{10.1117/12.2677095}

\bibitem[{{Potgieter}(2013)}]{2013LRSP...10....3P}
{Potgieter}, M.~S. 2013, Living Reviews in Solar Physics, 10, 3,
  \dodoi{10.12942/lrsp-2013-3}

\bibitem[{{Reynolds} {et~al.}(2023){Reynolds}, {Kara}, {Mushotsky}, \&
  {Ptak}}]{Reynoldsetal2023}
{Reynolds}, C., {Kara}, E., {Mushotsky}, R.~F., \& {Ptak}, A. 2023, in Society
  of Photo-Optical Instrumentation Engineers (SPIE) Conference Series, Vol.
  12678, Optics + Photonics 2023: UV, X-Ray, and Gamma-Ray Space
  Instrumentation for Astronomy XXIII, 12678--49

\bibitem[{{Richardson}(2004)}]{2004SSRv..111..267R}
{Richardson}, I.~G. 2004, \ssr, 111, 267,
  \dodoi{10.1023/B:SPAC.0000032689.52830.3e}

\bibitem[{{Richardson}(2018)}]{2018LRSP...15....1R}
---. 2018, Living Reviews in Solar Physics, 15, 1,
  \dodoi{10.1007/s41116-017-0011-z}

\bibitem[{{Sabbah}(2000)}]{2000GeoRL..27.1823S}
{Sabbah}, I. 2000, \grl, 27, 1823, \dodoi{10.1029/2000GL003760}

\bibitem[{{Sabbah} \& {Kudela}(2011)}]{2011JGRA..116.4103S}
{Sabbah}, I., \& {Kudela}, K. 2011, Journal of Geophysical Research (Space
  Physics), 116, A04103, \dodoi{10.1029/2010JA015922}

\bibitem[{{Sarkar} {et~al.}(2021){Sarkar}, {Su}, {Randall}, {Gastaldello},
  {Trierweiler}, {White}, {Kraft}, \& {Miller}}]{2021MNRAS.501.3767S}
{Sarkar}, A., {Su}, Y., {Randall}, S., {et~al.} 2021, \mnras, 501, 3767,
  \dodoi{10.1093/mnras/staa3858}

\bibitem[{{Sarkar} {et~al.}(2022{\natexlab{a}}){Sarkar}, {Su}, {Truong},
  {Randall}, {Mernier}, {Gastaldello}, {Biffi}, \&
  {Kraft}}]{2022MNRAS.516.3068S}
{Sarkar}, A., {Su}, Y., {Truong}, N., {et~al.} 2022{\natexlab{a}}, \mnras, 516,
  3068, \dodoi{10.1093/mnras/stac2416}

\bibitem[{{Sarkar} {et~al.}(2022{\natexlab{b}}){Sarkar}, {Randall}, {Su},
  {Alvarez}, {Sarazin}, {Nulsen}, {Blanton}, {Forman}, {Jones}, {Bulbul},
  {Zuhone}, {Andrade-Santos}, {Johnson}, \&
  {Chakraborty}}]{2022ApJ...935L..23S}
{Sarkar}, A., {Randall}, S., {Su}, Y., {et~al.} 2022{\natexlab{b}}, \apjl, 935,
  L23, \dodoi{10.3847/2041-8213/ac86d4}

\bibitem[{{Sarkar} {et~al.}(2023){Sarkar}, {Randall}, {Su}, {Alvarez},
  {Sarazin}, {Jones}, {Blanton}, {Nulsen}, {Chakraborty}, {Bulbul}, {Zuhone},
  {Andrade-Santos}, \& {Johnson}}]{2023ApJ...944..132S}
---. 2023, \apj, 944, 132, \dodoi{10.3847/1538-4357/acae9f}

\bibitem[{{Sarkar} {et~al.}(2024){Sarkar}, {Andrade-Santos}, {van Weeren},
  {Kraft}, {Hoang}, {Shimwell}, {Nulsen}, {Foreman}, {Randall}, {Su},
  {Chakraborty}, {Jones}, {Miller}, {Bautz}, \& {Grant}}]{2024ApJ...962..161S}
{Sarkar}, A., {Andrade-Santos}, F., {van Weeren}, R.~J., {et~al.} 2024, \apj,
  962, 161, \dodoi{10.3847/1538-4357/ad1aac}

\bibitem[{{Str{\"u}der} {et~al.}(2001){Str{\"u}der}, {Briel}, {Dennerl},
  {Hartmann}, {Kendziorra}, {Meidinger}, {Pfeffermann}, {Reppin}, {Aschenbach},
  {Bornemann}, {Br{\"a}uninger}, {Burkert}, {Elender}, {Freyberg}, {Haberl},
  {Hartner}, {Heuschmann}, {Hippmann}, {Kastelic}, {Kemmer}, {Kettenring},
  {Kink}, {Krause}, {M{\"u}ller}, {Oppitz}, {Pietsch}, {Popp}, {Predehl},
  {Read}, {Stephan}, {St{\"o}tter}, {Tr{\"u}mper}, {Holl}, {Kemmer}, {Soltau},
  {St{\"o}tter}, {Weber}, {Weichert}, {von Zanthier}, {Carathanassis}, {Lutz},
  {Richter}, {Solc}, {B{\"o}ttcher}, {Kuster}, {Staubert}, {Abbey}, {Holland},
  {Turner}, {Balasini}, {Bignami}, {La Palombara}, {Villa}, {Buttler},
  {Gianini}, {Lain{\'e}}, {Lumb}, \& {Dhez}}]{2001A&A...365L..18S}
{Str{\"u}der}, L., {Briel}, U., {Dennerl}, K., {et~al.} 2001, \aap, 365, L18,
  \dodoi{10.1051/0004-6361:20000066}

\bibitem[{{Tomassetti}(2017)}]{2017arXiv171203178T}
{Tomassetti}, N. 2017, arXiv e-prints, arXiv:1712.03178,
  \dodoi{10.48550/arXiv.1712.03178}

\bibitem[{{Torrence} \& {Compo}(1998)}]{1998BAMS...79...61T}
{Torrence}, C., \& {Compo}, G.~P. 1998, Bulletin of the American Meteorological
  Society, 79, 61, \dodoi{10.1175/1520-0477(1998)079<0061:APGTWA>2.0.CO;2}

\bibitem[{{Usoskin}(2017)}]{2017LRSP...14....3U}
{Usoskin}, I.~G. 2017, Living Reviews in Solar Physics, 14, 3,
  \dodoi{10.1007/s41116-017-0006-9}

\bibitem[{{von Kienlin} {et~al.}(2018){von Kienlin}, {Eraerds}, {Bulbul},
  {Fioretti}, {Gastaldello}, {Grant}, {Hall}, {Holland}, {Keelan}, {Meidinger},
  {Molendi}, {Perinati}, \& {Rau}}]{2018SPIE10699E..1IV}
{von Kienlin}, A., {Eraerds}, T., {Bulbul}, E., {et~al.} 2018, in Society of
  Photo-Optical Instrumentation Engineers (SPIE) Conference Series, Vol. 10699,
  Space Telescopes and Instrumentation 2018: Ultraviolet to Gamma Ray, ed.
  J.-W.~A. {den Herder}, S.~{Nikzad}, \& K.~{Nakazawa}, 106991I,
  \dodoi{10.1117/12.2311987}

\bibitem[{{Weisskopf} {et~al.}(2000){Weisskopf}, {Tananbaum}, {Van Speybroeck},
  \& {O'Dell}}]{2000SPIE.4012....2W}
{Weisskopf}, M.~C., {Tananbaum}, H.~D., {Van Speybroeck}, L.~P., \& {O'Dell},
  S.~L. 2000, in Society of Photo-Optical Instrumentation Engineers (SPIE)
  Conference Series, Vol. 4012, X-Ray Optics, Instruments, and Missions III,
  ed. J.~E. {Truemper} \& B.~{Aschenbach}, 2--16, \dodoi{10.1117/12.391545}

\end{thebibliography}
\bibliographystyle{aasjournal}







\label{lastpage}

\end{document}